\documentclass[fleqn,usenatbib]{mnras}

\usepackage[T1]{fontenc}
\usepackage{ae,aecompl}
\usepackage{comment}

\usepackage{graphicx}	
\usepackage{amsmath}	
\usepackage{amssymb}	
\usepackage[off]{auto-pst-pdf}
\usepackage{lineno}
\usepackage{epsfig}   
\usepackage{natbib}  
\usepackage{endnotes}
\usepackage{color}
\usepackage{fixfoot}
\usepackage{booktabs}

\title[Comptonizing medium of MAXI~J1535$-$571]{The comptonizing medium of the black-hole X-ray binary MAXI~J1535$-$571 through type-C quasi-periodic oscillations}

\author[Rawat et. al.]{Divya Rawat$^{1}$\thanks{E-mail: rawatdivya838@gmail.com (DR)},
Mariano M\'endez$^{2}$,
Federico Garc\'ia$^{2,3,4}$,
Diego Altamirano$^{5}$,
Konstantinos Karpouzas$^{2,5}$,
\newauthor Liang Zhang$^{5}$,
Kevin Alabarta$^{2,5}$,
Tomaso M. Belloni$^{6}$,
Pankaj Jain$^{7}$,
Candela Bellavita$^{4}$
\\
$^{1}$Inter-University Center for Astronomy and Astrophysics, Ganeshkhind, Pune 411007, India\\
$^{2}$Kapteyn Astronomical Institute, University of Groningen, PO BOX 800, Groningen NL-9700 AV, the Netherlands\\
$^{3}$Instituto Argentino de Radioastronom\'ia (CCT La Plata, CONICET; CICPBA; UNLP), C.C.5, (1894) Villa Elisa, Buenos Aires, Argentina\\
$^{4}$Facultad de Ciencias Astron\'omicas y Geof\'{\i}sicas, Universidad Nacional de La Plata, Paseo del Bosque, B1900FWA La Plata, Argentina\\
$^{5}$School of Physics and Astronomy, University of Southampton, Southampton SO17 1BJ, UK\\
$^{6}$INAF-Osservatorio Astronomico di Brera, via E. Bianchi 46, I-23807, Merate, Italy\\
$^{7}$Department of physics, IIT Kanpur, Kanpur, Uttar Pradesh 208016, India\\
}

\date{Accepted XXX. Received YYY; in original form ZZZ}

\begin{document}
\label{firstpage}
\pagerange{\pageref{firstpage}--\pageref{lastpage}}
\maketitle

\begin{abstract}
We present a detailed spectral and temporal analysis of the black-hole candidate MAXI~J1535$-$571 using NICER observations in September and October 2017. We focus specifically on observations in the hard-intermediate state when the source shows type-C quasi-periodic oscillations (QPOs). 
We fitted the time-averaged spectrum of the source and the rms and phase-lag spectra of the QPO with a one-component time-dependent Comptonization model. We found that the corona contracts from $\sim 10^4$ to $\sim 3 \times 10^3$ km as the QPO frequency increases from $\sim 1.8$ Hz to $\sim 9.0$ Hz. The fits suggest that the system would consists of two coronas, a small one that dominates the time-averaged spectrum and a larger one, possibly the jet, that dominates the rms and lag spectra of the QPO. We found a significant break in the relation of the spectral parameters of the source and the properties of the QPO, including its lag spectra, with QPO frequency. The change in the relations happens when the QPO frequency crosses a critical frequency $\nu_c \approx 3.0$ Hz. Interestingly, the QPO reaches this critical frequency simultaneously as the radio emission from the jet in this source is quenched.
\end{abstract}



\section{Introduction}
In the outburst, the transient black-hole X-ray binary (BHXB) system shows substantial X-ray variability \citep{be14}. These systems spend long periods in quiescence, with sporadic outbursts lasting weeks to months, during which the X-ray flux increases by up to three orders of magnitude compared to the quiescent phase \citep{re06}. During an outburst, transient BHXBs initially appear in the low-hard state (LHS) and, as the outburst progresses, move to the high-soft state (HSS) via the hard-intermediate (HIMS) and soft-intermediate state (SIMS) \citep[][and references within]{be05,be11}. Finally, before returning to the quiescent state, BHXBs transition from the HSS to the LHS. In the LHS, a hard component due to Comptonization from an electron plasma with temperature $50-100$ keV appears in the X-ray spectrum as a power law with photon index 1.5--2.0 \citep{gi10}. In contrast, the HSS spectrum is dominated by an optically thick thermal component generally modelled with a multi-temperature disc blackbody, occasionally accompanied by a soft power-law-like component with $\Gamma \ge$2 \citep{me97,do07}. The evolution of the outburst of a BHXB can be best characterised in a hardness-intensity diagram (HID), where typically systems trace a well-defined path often shaped as a ``q" (\citealt{fe04}, \citealt{be05}).\\

These systems show complex fast-time variability, which is strongly state-dependent. This variability takes the form of broadband noise components on top of which, in specific states, quasi-periodic oscillations (QPOs) can be observed (e.g. \citealt{ch97,ta97,ps99,no2000,ca04,ca05,be05}). The QPOs appear in the power density spectrum \citep[PDS;][]{va85} as relatively narrow peaks. The QPOs have been broadly divided into three categories, the mHz QPO with QPO frequency ranging from few mHz to Hz (e.g., \citealt{de06}, \citealt{ko11}, \citealt{al12}, \citealt{pa13}), low-frequency QPOs (LFQPOs) with frequencies ranging from just below 1 Hz up to 20 Hz \citep[e.g.,][]{mo15}, and high-frequency QPOs (HFQPOs) with frequencies above 100 Hz and up to $\sim$500 Hz (e.g., \citealt{mi01}, \citealt{st01}, \citealt{be12}, \citealt{me13}, \citealt{be14}). LFQPOs appear in different spectral states and have been further classified as type A, B, and C (\citealt{wi99}, \citealt{ho01}, \citealt{re02}, \citealt{ca04}). Among the three types, type-C is the one that is most often observed, showing a high rms amplitude, between 1$\%$ and 20$\%$, and a quality factor\footnote{\label{note1}{Quality factor=QPO frequency/QPO width}} usually larger than 6.0 (\citealt{wi99,ca04,be14}, see \citealt{in19}, for a review).\\

MAXI~J1535$-$571 (hereafter MAXI~J1535) is a galactic transient, initially detected by MAXI/GSC \citep{ne17a} and SWIFT/BAT (\citealt{ke17}, \citealt{mr17}) on September 2, 2017. The X-ray variability \citep{ne17b}, optical \citep{sc17} and near-infrared \citep{di17} properties of the source suggest that MAXI~J1535 is a low-mass X-ray binary (LMXB) source. Radio observations with the Australia Telescope Compact Array (ATCA) show a signature of a compact radio jet \citep{ru17}; this and the observed luminosity suggest that this system harbours a black hole \citep{ne17b}. Study of radio \citep{ca19} and X-ray \citep{sr19} observations suggest that the distance to the source is 4--6 kpc, and the jet inclination angle is constrained to $\le 45^\circ$ \citep{ru19}.
X-ray spectral studies suggest that the system harbours a near-maximally spinning black hole (\citealt{ge17a}, \citealt{xu18}, \citealt{mi18}). There are some conflicting estimates of the mass of the black hole in the system (\citealt{sh19}, \citealt{sr19}), but they are all based on fits to the X-ray spectrum and are therefore model dependent. No dynamical mass measurement from optical observations is available.\\

A state transition study of MAXI~J1535 during outburst, from September 2017 to April 2018 \citep{na18} shows that the source behaved like other BHXB systems tracing a q-shape in the HID \citep{ta18}. In the LHS and HIMS, starting from September 9-18, 2017, MAXI~J1535 showed a type-C QPO with a centroid frequency in the 0.2-3.4 Hz range (\citealt{ge17a}, \citealt{me18}, \citealt{st18}, \citealt{hu18}, \citealt{bh19}). The source transitioned to the SIMS and then to the HSS from September 19-26, 2017. The stable and weak type A/B LFQPO appears in the SIMS (\citealt{st18}, \citealt{se18}, \citealt{hu18}).  In the HIMS and LHS, the type-C QPO reappears from September 26 to October 9, 2017. After the end of the main outburst in mid-May 2018, five re-brightening events were reported by \citet{pa19}. A state transition during these re-flares was reported by \citet{cu20} using NICER observations.\\~\\ 

\citet{ku14} proposed a model to study the Comptonisation medium of neutron-star X-ray binary systems, which was later extended by \citet{ka20}. This model was originally developed for high-frequency QPOs in accreting neutron-star systems. Still, it has been recently extended by \citet{be22} to LFQPOs in BHXBs and was applied to the type-C QPO in GRS 1915+105 by \citet{ka21} and  \citet{me22}, and the type-B QPO in MAXI J1348$-$630 \citep{ga21,be22}. 
\citet{zh22} has applied the same model using Insight-HXMT observations of the type-C QPO in MAXI J1535  up to 150 keV.
The rationale behind applying this model to type-C in BHXB is that the fractional rms amplitude of these QPOs can be as large as $\sim$ 15\% up to $\sim$200 keV \citep{ma21}. At those energies, Comptonization dominates the emission in these systems (e.g., the disc and the reflection component peak at, respectively, $\sim$1$-$3 keV and $\sim$ 20$-$25 keV and both drop quickly above that), and hence Comptonization is most likely responsible for the rms amplitude and lags of the QPO.\\

In this paper, we report the results of the spectro-temporal analysis of MAXI~J1535 using NICER observations. To study the Comptonization medium of the source, we fit the rms and phase-lag spectra of the QPO with a one-component time-dependent Comptonization model, {\sc{vkompthdk}} \citep{ka20, be22}. In Section 2, we describe the observations and data analysis techniques, and in Section 3 we present the results of our analysis and the fits of the model to the rms and lag spectra of the type-C QPO. Finally, we discuss our findings in Section 4 and summarise our results in Section 5.

\begin{figure*}
\centering\includegraphics[width=0.48\textwidth,height=.225\textheight,angle=0]{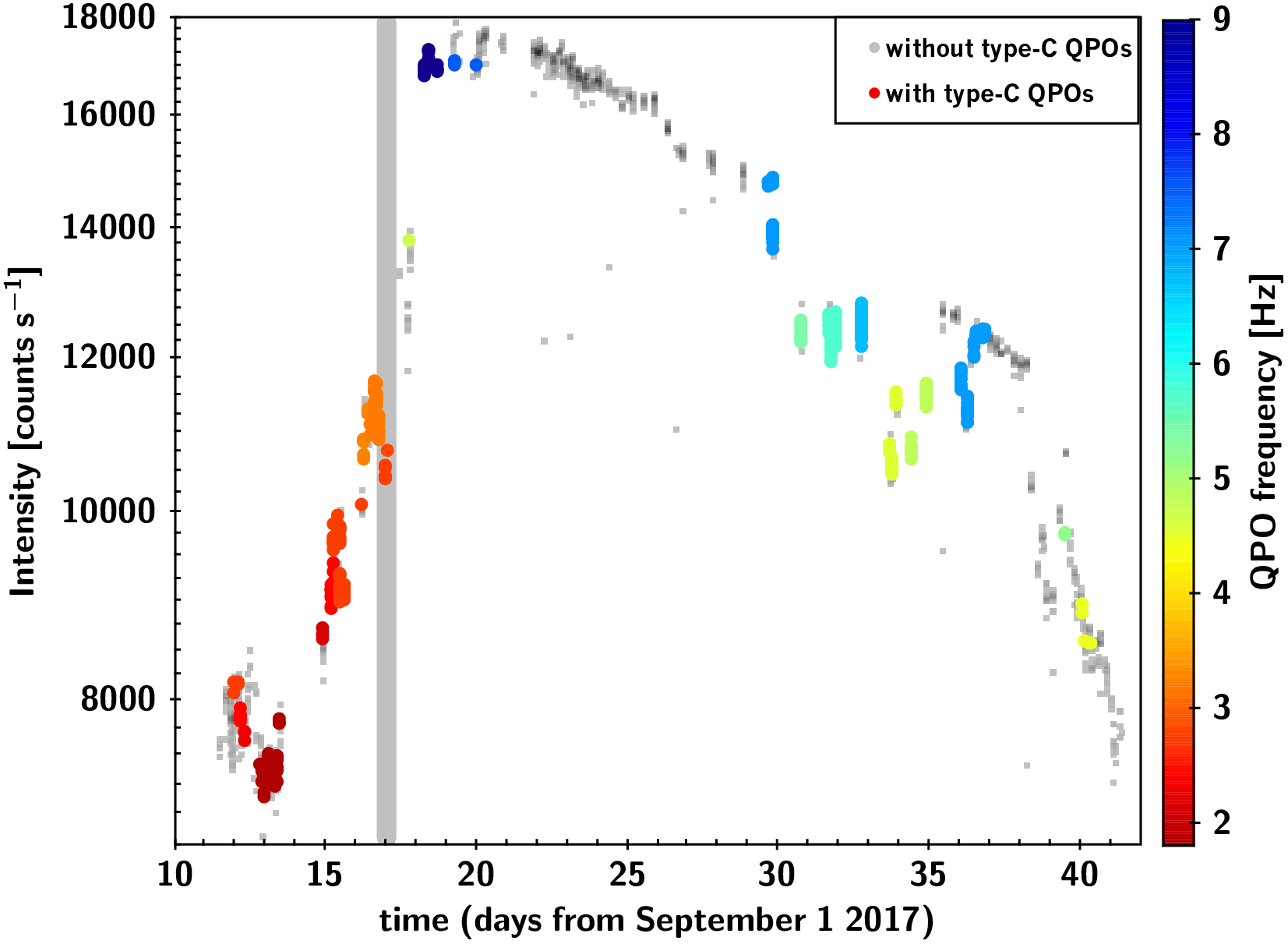}
\centering\includegraphics[width=0.48\textwidth, height=.225\textheight,angle=0]{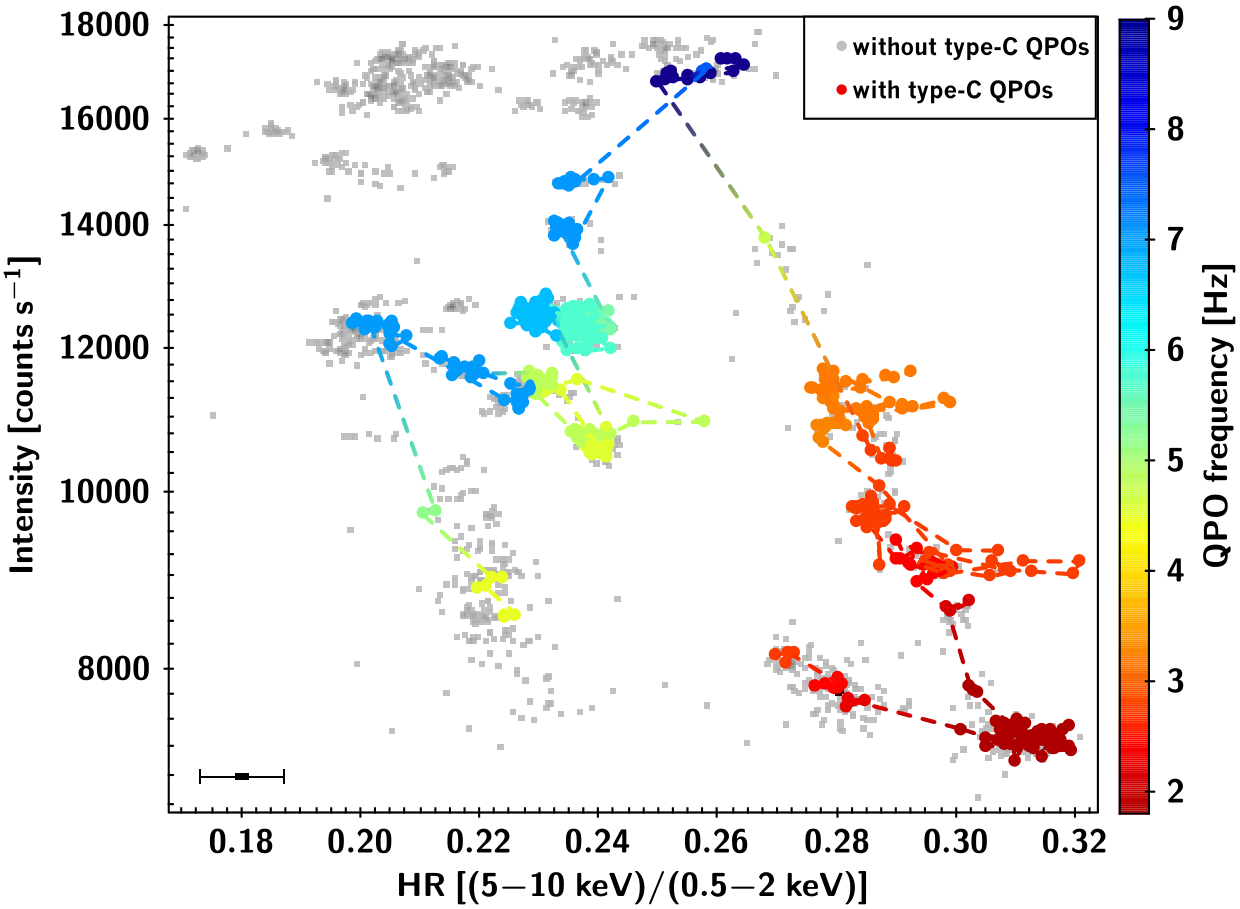}
\caption{Left panel: NICER light curve of MAXI~J1535$-$571 in the 0.5-10.0 keV band. The shaded area represents the approximate time when the radio emission was quenched \citep{ru19}. Right panel: Hardness intensity diagram (HID) using NICER observations. In the HID, the line shows the general movement of the source in this diagram as the outburst progressed, with the start and end points of the outburst at,  (HR $=0.27$, Intensity $=8000$) and (HR $=0.22$, Intensity $=8000$), respectively. In both panels, each point corresponds to 100 sec, and the colour scale panels indicate the frequency of the QPO. }
\label{HID}
\end{figure*}
\begin{table*}
 \centering
\caption{Observation log of MAXI J1535, including timing parameters. The columns are the observation number, the NICER ObsID, the start and end time of the observation, the 0.5-10.0 keV count rate, the standard deviation of the count rate, $\sigma_{count}$, the hardness ratio, HR, the standard deviation of the hardness ratio, $\sigma_{HR}$, the QPO centroid frequency and the QPO fractional rms amplitude. The errors are at 1$\sigma$. The observations with an asterisk are those for which the QPO was insignificant in the lowest energy bands.}
\begin{center}
\scalebox{0.95}{%
\begin{tabular}{cccccccccc}
\hline  
\hline
Obs no. &ObsID & Tstart & Tstop & count rate & $\sigma_{count}$ & HR &  $\sigma_{HR}$ & QPO frequency  & QPO Fractional  \\ & & (M.J.D) & (M.J.D) & (0.5-10.0 keV) & & $\frac{\rm({5-10 keV)}}{\rm{(0.5-2.0 keV)}}$ & & (Hz) & rms (\%)  \\ \hline 
1 &1050360105    & 58008.988  & 58009.126 & $8140\pm{5}$ & 48 &0.272 & 0.002& $2.74 \pm{0.01}$ & $7.0\pm{0.2}$ \\
2 &1050360105    & 58009.165  & 58009.193 & $7847\pm{4}$ & 36 & 0.280 & 0.002 &$2.44 \pm{0.01}$ & $6.5\pm{0.2}$ \\
3 &1050360105    & 58009.229  & 58009.301 & $7676\pm{6}$ & 30 & 0.285  & 0.004 &$2.32 \pm{0.01}$ & $6.7\pm{0.2}$ \\
4 &1050360105    & 58009.807  & 58009.945 & $7327\pm{4}$ & 65 & 0.307 & 0.003&$1.83 \pm{0.01}$ & $7.3\pm{0.2}$ \\
5 &1050360106    & 58010.001  & 58010.525 & $7364\pm{1}$ & 138 & 0.311 & 0.005& $1.81 \pm{0.00}$ & $7.2\pm{0.1}$ \\
6 &1050360107    & 58011.865  & 58011.940 & $8654\pm{7}$ & 47 & 0.299 & 0.002 &$2.15 \pm{0.01}$ & $6.9\pm{0.2}$ \\
7 &1050360108    & 58012.187  & 58012.258 & $9134\pm{3}$ & 130 & 0.294 & 0.006&$2.41 \pm{0.01}$ & $7.4\pm{0.2}$ \\
8 &1050360108    & 58012.316  & 58012.583 & $9492\pm{2}$ & 320 & 0.285  & 0.002 &$2.77 \pm{0.01}$ & $7.3\pm{0.2}$ \\
9 &1050360109    & 58013.216  & 58013.222 & $10088\pm{1}$ & 4 & 0.285 & 0.004 &$2.75 \pm{0.02}$ & $7.0\pm{0.2}$ \\
10 &1050360109    & 58013.281  & 58013.410 & $10922\pm{4}$ & 191 &0.275 & 0.008 & $3.27 \pm{0.02}$ & $7.0\pm{0.3}$ \\
11 &1050360109    & 58013.481  & 58013.740 & $11290\pm{2}$ & 227 & 0.282  & 0.005 &$3.19 \pm{0.03}$ & $6.7\pm{0.3}$ \\
12 &1050360109    & 58013.988  & 58013.998 & $10461\pm{5}$ & 71 & 0.288& 0.001 & $2.72 \pm{0.01}$ & $6.7\pm{0.2}$ \\
13 &1050360110    & 58014.053  & 58014.063 & $10744\pm{1}$ & 5 & 0.286 & 0.002 &$2.84 \pm{0.01}$ & $7.5\pm{0.2}$ \\
14 &1050360110    & 58014.824  & 58014.835 & $13795\pm{1}$ & 5 & 0.269 & 0.003 &$4.75 \pm{0.01}$ & $5.7\pm{0.1}$ \\
15 &1050360111    & 58015.276  & 58015.669 & $16992\pm{3}$ & 161 & 0.257 & 0.005 & $9.01 \pm{0.04}$ & $1.7\pm{0.1}$ \\
16 &\hspace{-0.15cm}$^{\ast}$1050360112    & 58016.240  & 58016.957 & $17040\pm{9}$ & 31 & 0.256 & 0.010 & $7.55 \pm{0.06}$ & $2.6\pm{0.2}$ \\
17 &1050360113    & 58017.011  & 58017.858 & $16995\pm{1}$ & 7 &0.244 & 0.017 & $7.45 \pm{0.03}$ & $2.9\pm{0.1}$ \\
18 &\hspace{-0.15cm}$^{\ast}$1130360103    & 58026.726  & 58026.814 & $14304\pm{2}$ & 445 & 0.235 & 0.002 & $7.09 \pm{0.03}$ & $2.4\pm{0.1}$ \\
19 &1130360104    & 58027.755  & 58027.779 & $12363\pm{3}$ & 105 & 0.240  & 0.002 &$5.42 \pm{0.01}$ & $4.7\pm{0.1}$ \\
20 &1130360105    & 58028.720  & 58028.872 & $12321\pm{2}$ & 213 &0.237 & 0.002 &$5.73 \pm{0.01}$ & $4.5\pm{0.0}$ \\
21 &\hspace{-0.15cm}$^{\ast}$1130360106    & 58029.749  & 58029.836 & $12527\pm{2}$ & 151 & 0.229& 0.002 & $6.77 \pm{0.02}$ & $3.5\pm{0.1}$ \\
22 &1130360107    & 58030.715  & 58030.865 & $10831\pm{2}$ & 381 & 0.238 & 0.004 &$4.57 \pm{0.01}$ & $4.6\pm{0.1}$ \\
23 &1130360108    & 58031.361  & 58031.894 & $11163\pm{2}$ & 370 & 0.234  & 0.006 & $4.82 \pm{0.01}$ & $3.5\pm{0.0}$ \\
24 &1130360113    & 58036.498  & 58036.695 & $9747\pm{10}$ & 19 & 0.206 & 0.007 &$5.19 \pm{0.03}$ & $3.0\pm{0.2}$ \\
25 &1130360114    & 58037.032  & 58037.677 & $8767\pm{4}$ & 183 & 0.224 &0.004 &$4.50 \pm{0.01}$ & $5.0\pm{0.1}$ \\
\hline
\end{tabular}}
\label{table1}
\end{center}
\end{table*}

\begin{table*}
 \centering
 \caption{Time-averaged spectra and corona model parameters of MAXI J1535. The columns are the observation number, the hydrogen column density, $N_{\rm H}$, the power-law photon index of {\sc{nthcomp}}, $\Gamma$, the inner disc temperature, $kT_{in}$, the seed photon temperature of {\sc{vkompthdk}}, $kT_{s}$, the size of the corona, $L$, the fraction of the flux of the seed-photon source due to feedback from the corona, $\eta$, and the amplitude of the variability of the external heating rate, $\delta \dot{H}_{ext}$. The errors are at 1$\sigma$. The observations with an asterisk are those for which the QPO was insignificant in the lowest energy bands.}
\begin{center}
\scalebox{1.0}{%
\hspace{-0.5cm}
\begin{tabular}{ccccccccc}
\hline  
\hline
 Obs no. & $N_{\rm H}$                   &  $\Gamma$                   & $kT_{in}$                    &     $kT_{s}$             &      $L$            &           $\eta$           &    $\delta \dot{H}_{ext}$        & $\chi^{2}_{\nu}$(dof) \\   & $10^{22}$ cm$^{-2}$ &  &(keV)   &(keV)  & ( $10^{3}$ km) &  &  \% &  \\ \hline 
1 & $2.19\pm{0.01}$         & $2.43\pm{0.02}$           & $0.68\pm{0.01}$            & $0.35 \pm 0.05$        & $5.1 \pm 1.0$         & $0.62\pm{0.05}$          &  $12.2 \pm 0.6$         & $231.4 \;(243)$ \\
2 & $2.19\pm{0.01}$         & $2.29\pm{0.01}$           & $0.62\pm{0.01}$            & $0.29 \pm 0.03$        & $8.3 \pm 1.1$         & $0.75\pm{0.09}$          &  $12.0 \pm 0.5$         & $191.9 \;(242)$ \\
3 &$2.18\pm{0.01}$         & $2.26\pm{0.01}$           & $0.61\pm{0.01}$            & $0.23 \pm 0.04$        & $8.7 \pm 1.1$         & $0.82^{+0.18}_{-0.38}$          &  $11.3 \pm 1.1$         & $240.5 \;(243)$ \\
4 &$2.17\pm{0.01}$         & $2.12\pm{0.01}$           & $0.55\pm{0.01}$            & $0.14 \pm 0.01$        & $12.6 \pm 0.5$         & $1.00-{0.04}$          &  $11.1 \pm 0.4$         & $219.8 \;(243)$ \\
5 &$2.16\pm{0.01}$         & $2.11\pm{0.00}$           & $0.55\pm{0.01}$            & $0.15 \pm 0.01$        & $13.2 \pm 0.4$         & $1.00-{0.45}$          &  $11.5 \pm 0.3$         & $242.3 \;(243)$ \\
6 &$2.15\pm{0.01}$         & $2.18\pm{0.01}$           & $0.60\pm{0.01}$            & $0.24 \pm 0.03$        & $9.1 \pm 1.0$         & $0.79\pm{0.12}$          &  $12.2 \pm 0.7$         & $177.9 \;(243)$ \\
7 &$2.17\pm{0.01}$         & $2.27\pm{0.01}$           & $0.64\pm{0.01}$            & $0.36 \pm 0.05$        & $6.6 \pm 1.2$         & $0.64\pm{0.07}$          &  $15.0 \pm 0.7$         & $173.2 \;(243)$ \\
8 &$2.15\pm{0.01}$         & $2.67\pm{0.04}$           & $0.79\pm{0.01}$            & $0.33 \pm 0.04$        & $5.7 \pm 0.9$         & $0.76\pm{0.07}$          &  $11.2 \pm 0.7$         & $234.8 \;(243)$ \\
9 &$2.19\pm{0.01}$         & $2.34\pm{0.02}$           & $0.68\pm{0.01}$            & $0.47 \pm 0.07$        & $4.8 \pm 0.9$         & $0.59\pm{0.05}$          &  $14.4 \pm 0.9$         & $169.3 \;(243)$ \\
10 &$2.21\pm{0.01}$         & $2.48\pm{0.02}$           & $0.74\pm{0.01}$            & $0.39 \pm 0.07$        & $4.4 \pm 1.2$         & $0.55\pm{0.07}$          &  $13.5 \pm 1.0$         & $155.1 \;(243)$ \\
11 &$2.19\pm{0.01}$         & $2.85\pm{0.11}$           & $0.85\pm{0.02}$            & $0.36 \pm 0.05$        & $5.5 \pm 1.2$         & $0.77\pm{0.11}$          &  $10.5 \pm 0.9$         & $192.2 \;(222)$ \\
12 &$2.20\pm{0.01}$         & $2.33\pm{0.01}$           & $0.67\pm{0.01}$            & $0.37 \pm 0.04$        & $6.5 \pm 0.8$         & $0.66\pm{0.05}$          &  $13.3 \pm 0.5$         & $152.4 \;(243)$ \\
13 &$2.20\pm{0.01}$         & $2.36\pm{0.01}$           & $0.69\pm{0.01}$            & $0.37 \pm 0.04$        & $6.4 \pm 0.8$         & $0.69\pm{0.06}$          &  $14.3 \pm 0.5$         & $176.3 \;(243)$ \\
14 &$2.23\pm{0.01}$         & $2.61\pm{0.06}$           & $0.98\pm{0.02}$            & $0.43 \pm 0.05$        & $3.8 \pm 0.5$         & $0.73\pm{0.06}$          &  $14.1 \pm 0.8$         & $168.4 \;(242)$ \\
15 &$2.30\pm{0.01}$         & $2.49\pm{0.17}$           & $1.18\pm{0.01}$            & $0.56 \pm 0.04$        & $4.0 \pm 0.5$         & $1.00-{0.11}$          &  $17.3 \pm 2.0$         & $195.8 \;(239)$ \\
16 &$2.29\pm{0.01}$         & $2.60\pm{0.14}$           & $1.13\pm{0.02}$            & $0.52 \pm 0.07$        & $3.7 \pm 1.0$         & $0.69\pm{0.18}$          &  $15.6 \pm 2.4$         & $165.0 \;(236)$ \\
17  &$2.29\pm{0.00}$         & $2.40\pm{0.24}$           & $1.19\pm{0.01}$            & $0.39 \pm 0.04$        & $3.4 \pm 0.2$         & $0.88\pm{0.04}$          &  $20.6 \pm 1.2$         & $177.0 \;(242)$ \\
18 &$2.27\pm{0.01}$         & $2.71\pm{0.10}$           & $1.05\pm{0.02}$            & $0.55 \pm 0.04$        & $3.0 \pm 0.3$         & $0.60\pm{0.06}$          &  $11.7 \pm 0.8$         & $244.4 \;(229)$ \\
19 &$2.24\pm{0.00}$         & $2.61\pm{0.08}$           & $0.95\pm{0.02}$            & $0.46 \pm 0.04$        & $3.8 \pm 0.5$         & $0.65\pm{0.05}$          &  $14.3 \pm 1.1$         & $203.6 \;(242)$ \\
20 &$2.30\pm{0.01}$         & $3.03\pm{0.02}$           & $0.85\pm{0.01}$            & $0.63 \pm 0.02$        & $4.0 \pm 0.2$         & $0.57\pm{0.03}$          &  $14.1 \pm 0.4$         & $204.5 \;(240)$ \\
21 &$2.31\pm{0.01}$         & $3.38\pm{0.05}$           & $0.92\pm{0.01}$            & $0.76 \pm 0.04$        & $2.7 \pm 0.2$         & $0.44\pm{0.03}$          &  $13.5 \pm 0.8$         & $198.2 \;(215)$ \\
22 &$2.31\pm{0.02}$         & $2.66\pm{0.03}$           & $0.71\pm{0.02}$            & $0.57 \pm 0.03$        & $4.8 \pm 0.4$         & $0.51\pm{0.04}$          &  $13.3 \pm 0.6$         & $258.0 \;(219)$ \\
23 &$2.28\pm{0.01}$         & $3.07\pm{0.04}$           & $0.85\pm{0.01}$            & $0.55 \pm 0.03$        & $4.5 \pm 0.4$         & $0.66\pm{0.05}$          &  $9.1 \pm 0.5$         & $223.1 \;(238)$ \\
24 &$2.21\pm{0.00}$         & $2.69\pm{0.08}$           & $0.96\pm{0.02}$            & $0.40 \pm 0.05$        & $6.2 \pm 1.5$         & $1.00-{0.33}$          &  $12.8 \pm 1.9$         & $191.4 \;(241)$ \\
25 &$2.23\pm{0.00}$         & $2.58\pm{0.03}$           & $0.82\pm{0.02}$            & $0.43 \pm 0.03$        & $5.5 \pm 0.7$         & $0.67\pm{0.08}$          &  $15.9 \pm 0.5$         & $174.9 \;(242)$ \\
\hline
\end{tabular}}
\end{center}
\label{table2}
\end{table*}
\section{Observation and Data Analysis}
We used observations of MAXI~J1535 obtained in September and October 2017 with the Neutron Star Interior Composition Explorer \citep[NICER][]{ge12}. The observations ID's used are 1050360101-1050360120 $\&$ 1130360101-1130360114. NICER's XTI \citep[X-ray Timing Instrument][]{ge16} covers the 0.2-12.0 keV band and has an effective area of $>$2000 cm$^{2}$ at 1.5 keV. The energy and time resolutions are 85 eV at 1 keV and 4 $\times 10^{-8}$ s (hereafter $\Delta t_{nicer}$), respectively. We used the {\sc{nicerl2}}\footnote{\url{https://heasarc.gsfc.nasa.gov/docs/nicer/analysis_threads/nicerl2/}} task to process each observation applying the standard calibration process and screening. We used only those intervals for which the exposure time was $>$ 100 s after running the {\sc{nicerl2}} task. For some intervals, we found that the source flux was varying significantly. To make sure we are not averaging features of two spectrally and temporally different states, we divided a single observation into segments with a more or less constant source count rate and studied the temporal and spectral properties of each segment independently. The details of each observation and segment are given in Table \ref{table1}. 
\subsection{Timing analysis}
We extracted the fractional rms amplitude (root-mean square) normalised \citep{be90} PDS for each segment using the General High-energy Aperiodic Timing Software (GHATS)\footnote{\url{http://www.brera.inaf.it/utenti/belloni/GHATS_Package/Home.html}} version 2.1.0. The 0.2-10.0 keV data were re-binned in time by a factor of 62500, such that the time resolution was 0.0025 s, corresponding to a Nyquist frequency of 200 Hz, and PDS were produced from intervals of 8192 points (20.48 s). For each segment, the PDS for the intervals were averaged. We fitted the PDS in the frequency 100-200 Hz, where the source shows no intrinsic variability, with a constant representing the Poisson noise, which we then subtracted. We ended up with an averaged, Poisson-noise subtracted PDS for each segment that we re-binned logarithmically such that each frequency bin is larger than the previous one by a factor exp(1/100). We fitted all the PDS with a model consisting of up to five Lorentzians to represent the broadband noise component and the QPOs. Each Lorentzian has three parameters: the centroid frequency, $\nu_0$, the full-width at half-maximum, FWHM, and the total power, equal to the integral of the Lorentzian function over the full frequency range. We only included a Lorentzian in the model if its total power was at least $3 \sigma$ different from zero, given the error of this parameter. We visually inspected the PDS from all segments and used only those with a clear type-C QPO.\\

Next, we extracted PDS in 10 energy bands, 1.0--1.5, 1.5--1.9, 1.9--2.3, 2.3--3.0, 3.0--3.5, 3.5--4.0, 4.0--5.0, 5.0--6.0, 6.0--8.0, and 8.0--12.0 keV that we normalised to fractional rms for each band. To extract phase/time lags, we computed FFTs from the data in the ten energy bands and measured the lags using the phases of the cross-spectra with the 2.0--3.0 keV band as a reference, following the procedure of \citet{no99}. To calculate the lags of the QPO, we averaged the cross spectra within one full-width half-maximum around the centroid frequency of the QPO for each segment in which we detected a significant QPO. For 4 segments, marked with an asterisk in Table \ref{table1}, the QPO was insignificant in the lowest energy bands. We merged some low-energy bands in those cases and extracted the rms and lag spectra for 7 energy bands (1.0--2.3, 2.3--3.5, 3.5--4.0, 4.0--5.0, 5.0--6.0, 6.0--8.0, and 8.0--12.0 keV).
\subsection{Spectral analysis}
We produced the spectra and background files using the NICER background estimator tool {\sc{3C 50\_RGv5}}\footnote{\url{https://heasarc.gsfc.nasa.gov/docs/nicer/tools/nicer_bkg_est_tools.html}}. The background-subtracted spectrum for each segment was re-binned using {\sc{grppha}} such that each spectral bin had at least 30 counts and the bins over-sampled the spectral resolution of the detector by a factor 3. We used {\sc{Heasoft}} version 6.30 and CALDB version 20210707 to create the response (rmf) and ancillary response (arf) files. We fitted the time-averaged spectrum of the source in the $1.0-10.0$ keV band using the model  {\sc tbabs*(diskbb+gauss+nthcomp)}
 in {\sc{xspec}}. The {\sc{Tbabs}} models the interstellar absorption. We used the cross-section tables of \citet{ve96} and the abundances of \citet{wi00} and left the hydrogen column density as a free parameter. The {\sc{diskbb}} component models the thermal emission from an optically thick and geometrically thin accretion disc (\citealt{mi84}, \citealt{ma86}) while {\sc{nthcomp}} (\citealt{zd96},\citealt{zy99}) models the Comptonised emission from the X-ray corona. We kept both the {\sc{diskbb}} parameters, the temperature at inner disk radius, $kT_{in}$, and the normalisation free. The  {\sc{nthcomp}} model parameters are the power-law photon index, $\Gamma$, electron temperature, $kT_e$, seed photon temperature,  $kT_{bb}$, and normalization. The seed-photon temperature $kT_{bb}$ was tied to $kT_{in}$ of the {\sc{diskbb}} component. We have fitted a relatively broad iron line present in the residuals with a Gaussian, {\sc {gauss}} in {\sc{xspec}}. In addition to the broad line, the spectra show narrow residuals at $\sim$6.4 keV. We have added one more {\sc {gauss}} component to account for the narrow line (if required).\\

We fit the rms with the model {\sc{vkompthdk*dilution}}\footnote{\url{https://github.com/candebellavita/vkompth}} \citep{ka20,be22} and the lag spectra with the model {\sc vkompthdk} at the QPO frequency. {\sc vkompthdk} can compute both the time-dependent and the time-averaged spectrum. The time-dependent version of {\sc vkompthdk} is the one that fits the rms and lags. The time-averaged version of {\sc vkompthdk} is the same as {\sc{nthcomp}}. The parameters of {\sc vkompthdk} are hence the temperature of the seed photon source, $kT_s$, the temperature of the corona, $kT_e$, the power-law index, $\Gamma$ (all of them identical to $kT_{bb}$, $kT_e$ and $\Gamma$ of {\sc{nthcomp}}), plus the size of the corona, $L$, the feedback fraction, $\eta$ (between 0 to 1), the amplitude of the variability of the external heating rate, $\delta \dot{H}_{ext}$, and the lag of the model in the 2--3 keV energy band, {\rm{reflag}}. These parameters can be used to compute the fraction of the corona flux, $\eta_{int}$, that returns to the disc (see \citealt{ka20} for details). The parameters $L$, $\eta$, $\delta \dot{H}_{ext}$, and reflag are only relevant for the fits to the rms and lag spectra and do not affect the time-averaged version of the {\sc vkompthdk}. The parameter {\it{reflag}} is an additive normalisation that allows the model to match the data, given that the observer is free to choose the reference energy band of the lags. We froze the electron temperature of {\sc{nthcomp}} and {\sc vkompthdk} at $kT_e=21$ keV \citep{sr19} because the 1.0-10.0 keV energy band is not suitable to constrain it. The {\sc dilution} component is a function of energy (E). It accounts for the fact that the rms amplitude we observe is diluted by the emission of the other components that we assume do not vary. The {\sc{dilution}} component is therefore defined as;

 \begin{equation*}
  \rm{dilution(E)}= \frac{\rm{nthcomp(E)}}{\rm{diskbb(E)+gauss(E)+nthcomp(E)}}
 \end{equation*}
 (See details in \cite{be22}.) Because $N_{\rm H}$ towards the source is high, any emission below 1 keV could be attributed to calibration artefacts; therefore, we have decided to exclude data below 1.0 keV in our fits. Using HXMT data in the  2--100 keV range,  \citet{zh22} reported a hydrogen column density, $N_{\rm H}$=5.6*$10^{22}$ cm$^{-2}$, that is higher than the value we have obtained here using NICER in the 1--10 keV range.
\section{Results}
The left panel of Figure \ref{HID} shows the NICER light curve of MAXI~J1535 during its 2017 outburst. While the right panel of Figure \ref{HID} shows the evolution of the source in the HID. Here intensity is defined as the source count rate in the 0.5--10.0 keV band, and hardness ratio (HR) is the ratio of the source intensity in the  5.0--10.0 keV and 0.5--2.0 keV bands. The colour scale shown at the right of both Figures represents the QPO frequency range 1.8--9.0 Hz, with red being the lowest and navy blue being the highest end of the QPO frequency range. The source's X-ray count rate and HR and their respective standard deviation values for each segment are given in Table \ref{table1}. 
\\

\subsection{Spectral fits}
\begin{figure*}
\centering\includegraphics[scale=0.55,angle=0]{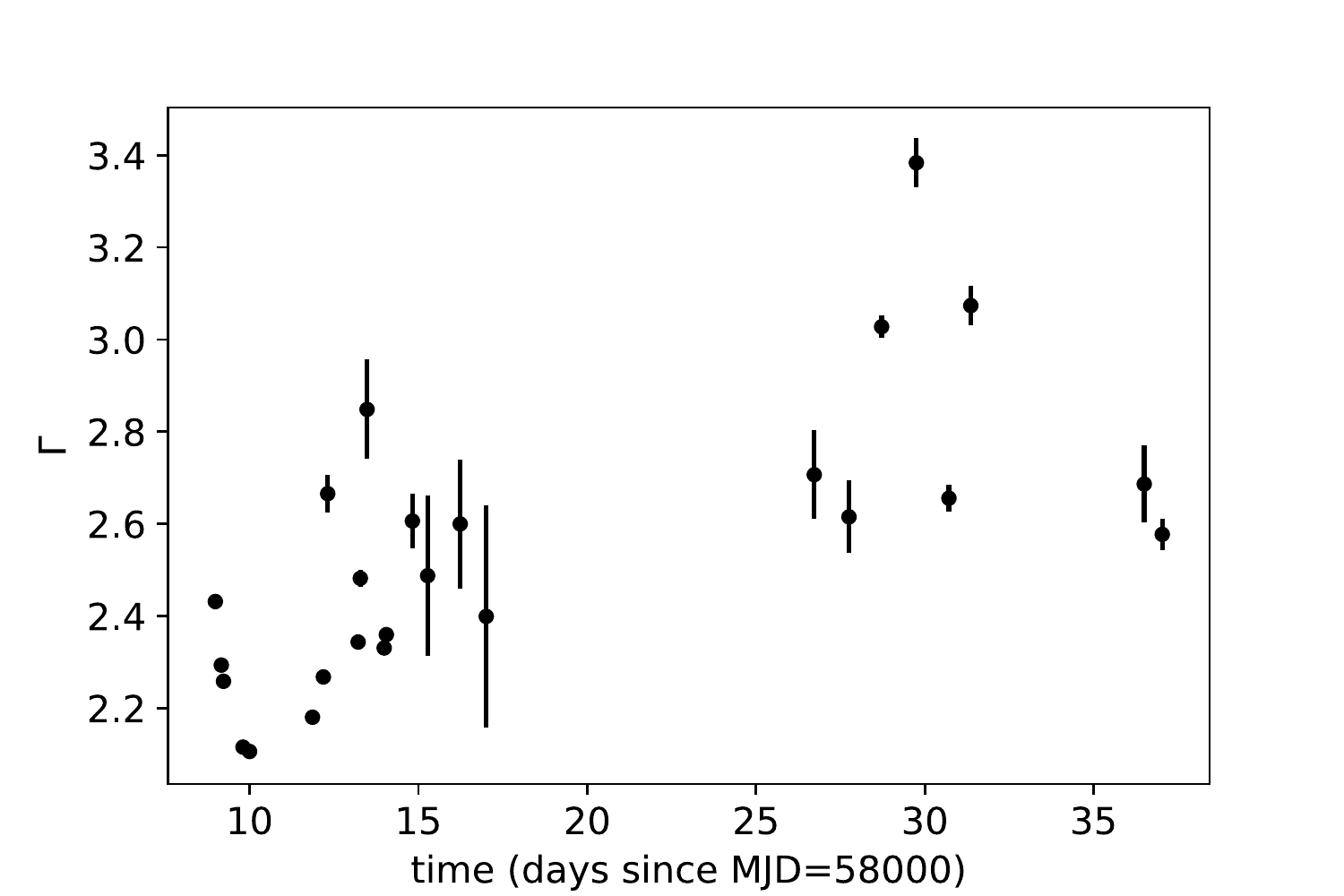}
\centering\includegraphics[scale=0.55,angle=0]{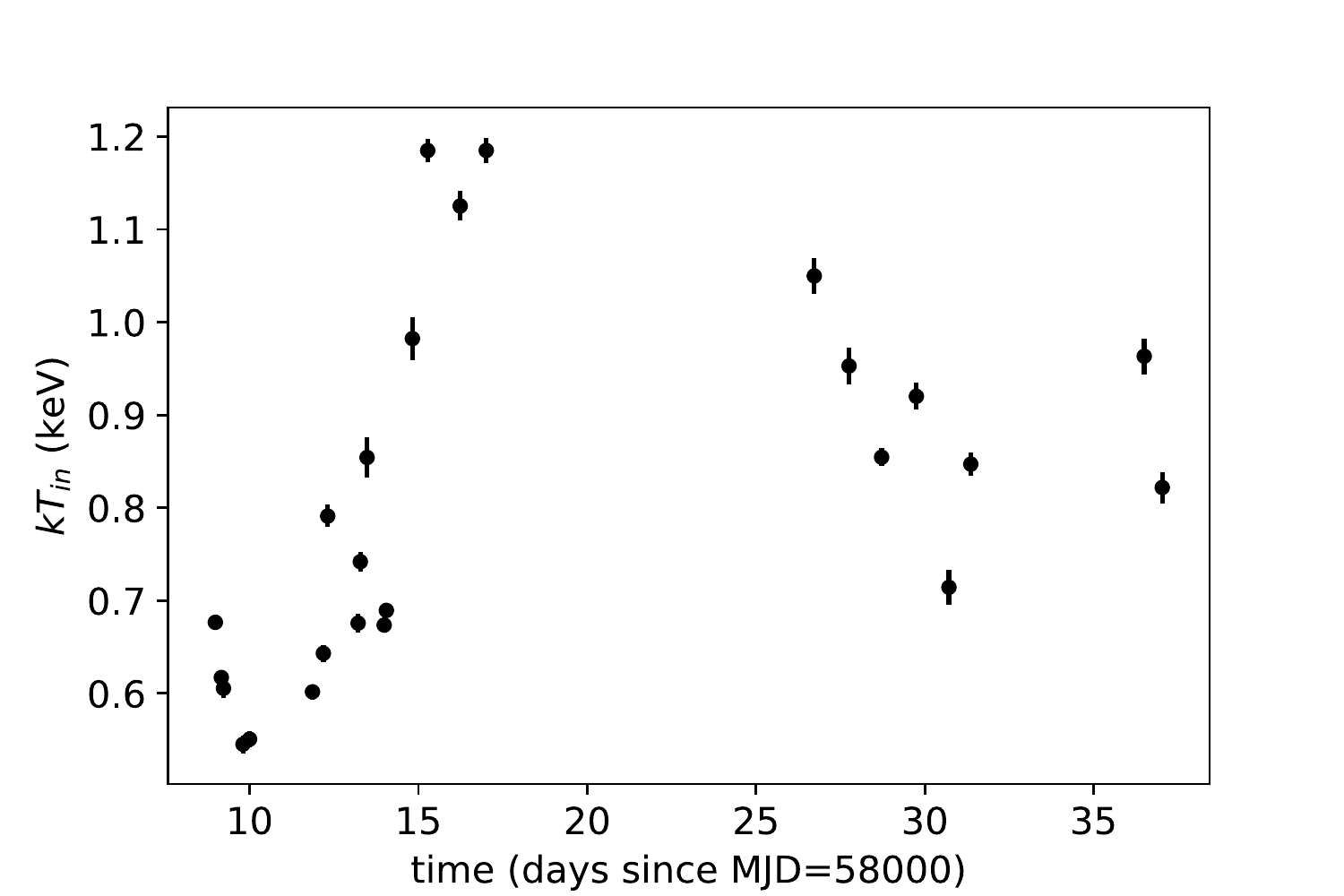}
\caption{The evolution of $\Gamma$ of the corona (left panel) and $kT_{in}$ of the disc (right panel) of MAXI J1535$-$571. The values of $\Gamma$ and $kT_{in}$ are obtained from the fits to the time-averaged spectra, the rms and phase-lag spectra of the QPO.}
\label{time_gamma}
\end{figure*}

From the fits to the time-averaged spectrum, the rms and phase-lag spectra of the QPO for each segment, we find that during the first two days of our observations, the inner disc temperature, $kT_{in}$, and the photon index, $\Gamma$, of the Comptonised component first drop (Figure \ref{time_gamma}) as the source moves to the right in the HID (Figure \ref{HID} right panel), from hardness ratio $\sim 0.27$ to hardness ratio $\sim 0.31$. Between MJD 58010 and MJD 58012, the source intensity increases, and the spectrum softens again. The source starts to move up and to the left in the HID, and $kT_{in}$ and $\Gamma$ increase very quickly for about five days. At the end of this period, the source reaches the highest intensity in our observations. The accretion disc is the hottest, $kT_{in} \approx 1.1-1.2$ keV, and the Comptonised component is described with $\Gamma \approx 2.7-2.8$. At this point, the source enters the HSS and the PDS show no QPOs. 
When the source transitions back to the SIMS and the HIMS, at around MJD  58025, $kT_{in}$ and $\Gamma$ are approximately correlated with the X-ray flux (see Figures \ref{HID} and \ref{time_gamma}).
We give each segment's spectral parameters and goodness of fit in Table \ref{table2}. In a few segments the reduced $\chi^2$ is less than 1 (last column of Table \ref{table2}). The low $\chi^2$ values come from the fit to the steady-state spectra (SSS). We provide the $\chi^2$ and the number of channels for the fits to the individual spectra and the total $\chi^2$ and the number of degree of freedom in Table \ref{table_appendix0}.
Unless otherwise specified, the errors represent the $1 \sigma$ confidence (68\%) interval for the corresponding parameter.

\begin{figure*}
\centering\includegraphics[width=1.0\textwidth,height=0.31\textheight,angle=0]{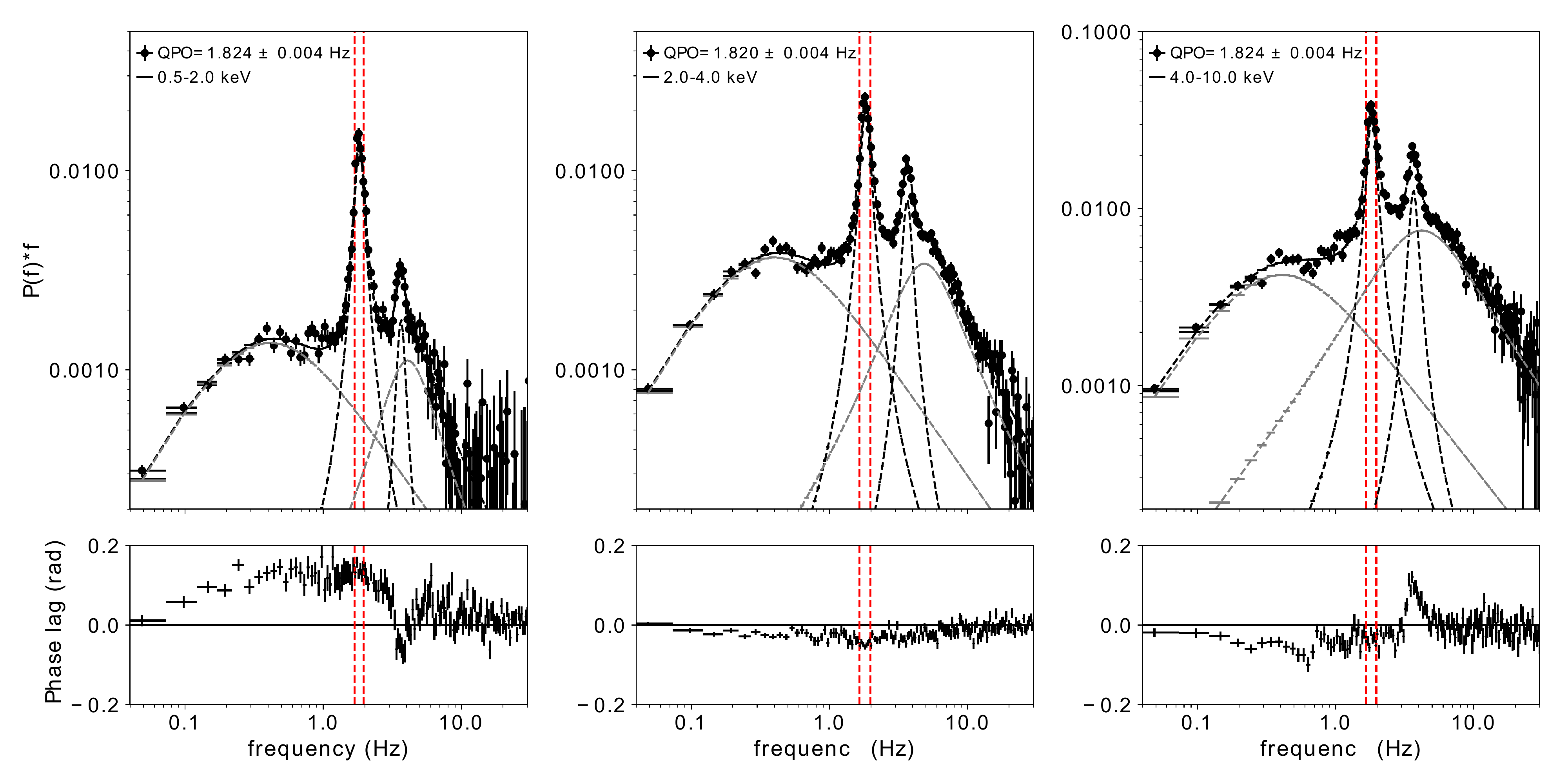}
\centering\includegraphics[width=1.0\textwidth,height=0.31\textheight,angle=0]{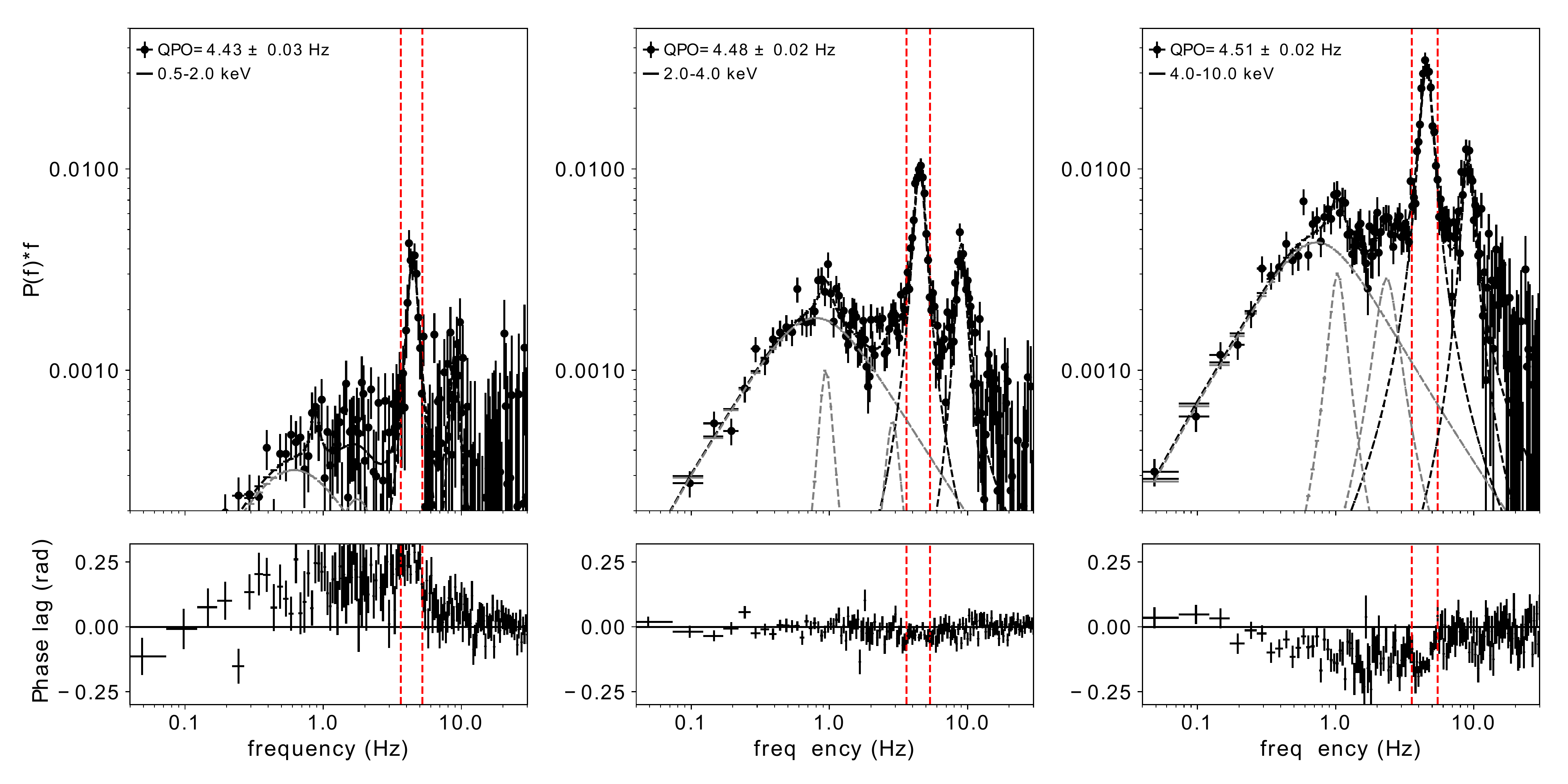}
\centering\includegraphics[width=1.0\textwidth,height=0.31\textheight,angle=0]{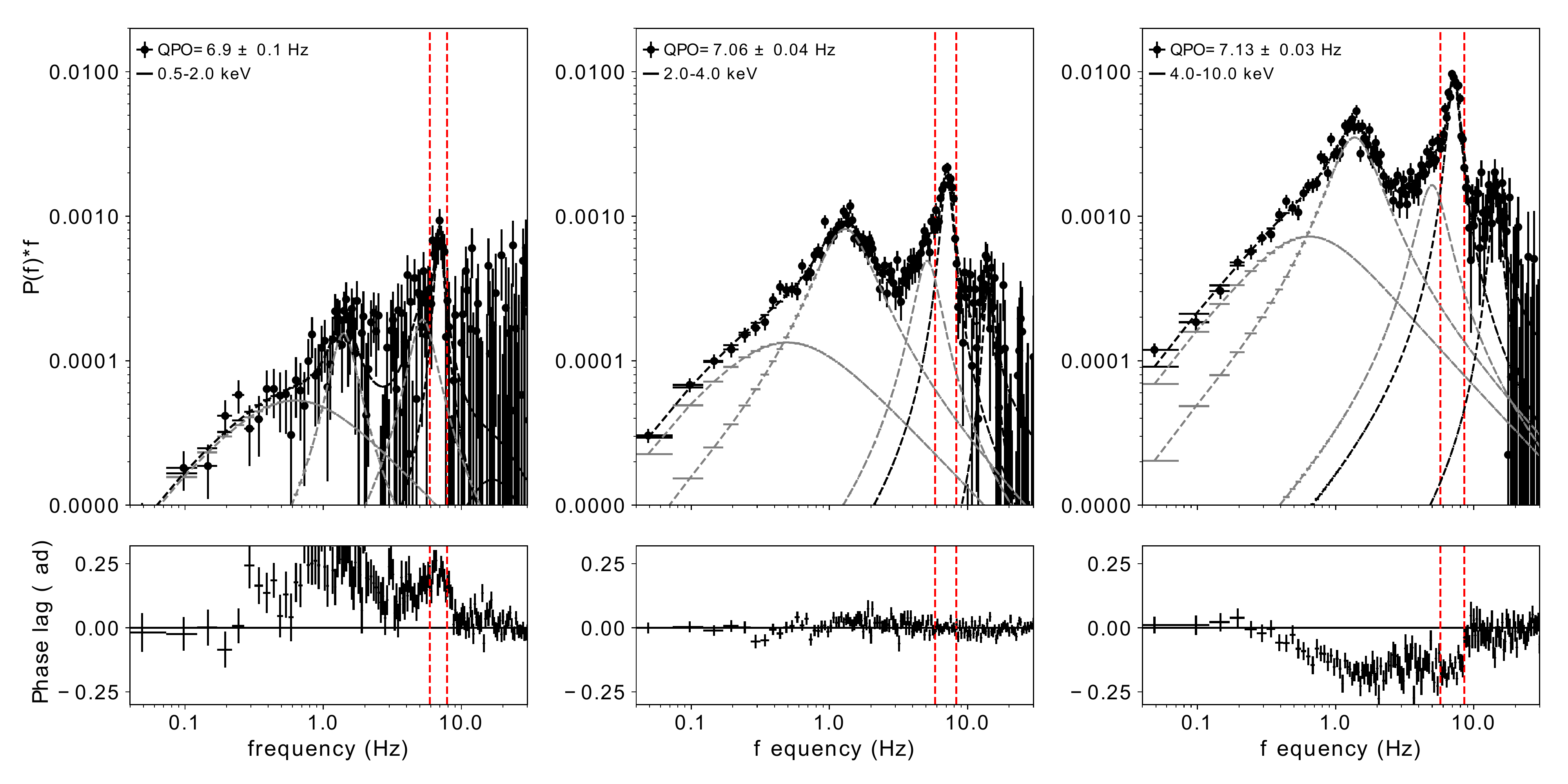}
\caption{The top panels show the power density spectra (power multiplied by frequency) of MAXI~J1535$-$571 for three QPO frequencies, 1.8 Hz, 4.5 Hz, and 7.0 Hz, and three different energy bands. The PDS is fitted with three to five Lorentzians. The bottom panels show the frequency phase-lag spectra. The reference energy band is 0.5-10.0 keV here. The vertical dashed lines indicate the ranges over which the QPO fundamental lags we measured ($\nu$ $\pm$ FWHM/2).}
\label{pds}
\end{figure*}
 \begin{figure*}
\centering\includegraphics[scale=0.55,angle=90]{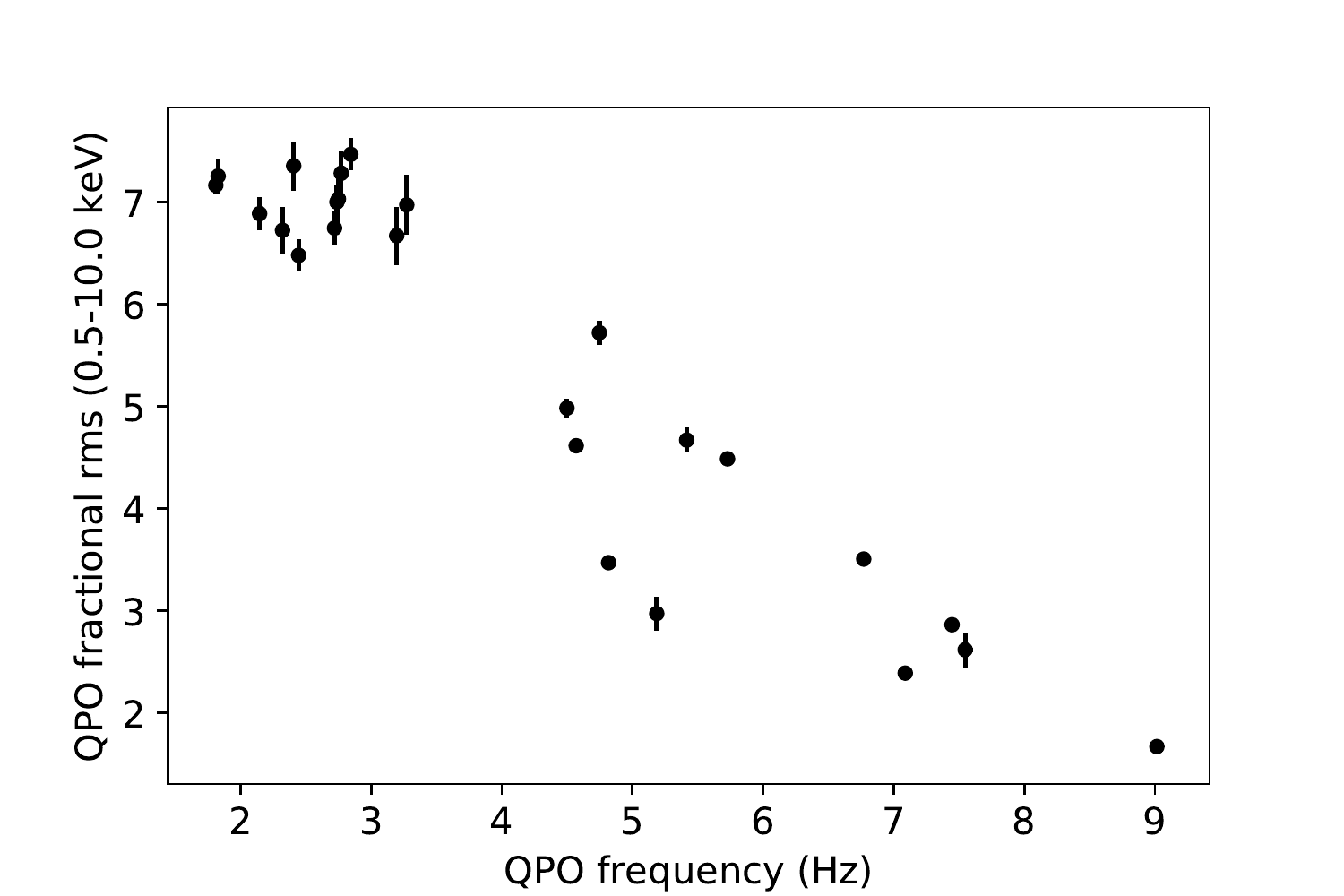}
\centering\includegraphics[scale=0.55,angle=0]{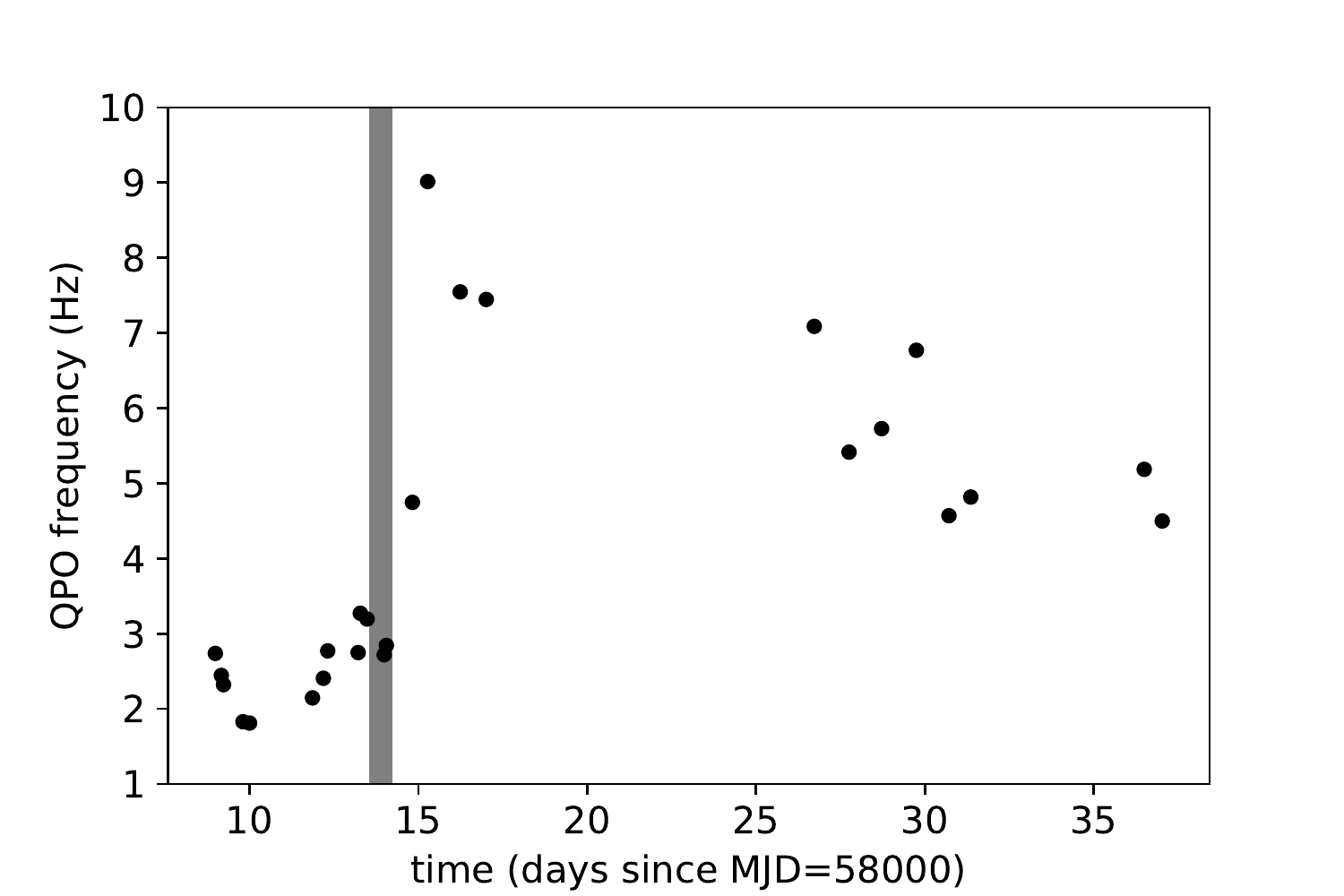}
\caption{Left panel: QPO fractional rms amplitude in the 0.5--10.0 keV energy band as a function of QPO frequency for MAXI J1535$-$571. Right panel: Evolution of the QPO frequency of MAXI 1535-571. The shaded area represents the radio jet quenching interval \citep{ru19}. }
\label{qpo_evolution}
\end{figure*}
\subsection{Power Density Spectra}
 Following \citet{be02}, we fit the PDS with a 0-centred Lorentzian to represent the broadband noise component and three separate Lorentzians to fit the narrow QPO, its harmonic component, and the high-frequency noise. The features in the PDS have a frequency in the ratio of 1:2, and we, therefore, identify the strongest peak as the fundamental and the other as the second harmonic. The PDS also shows a low-frequency noise component when the strongest QPO peak was at a frequency above 4.0 Hz (Figure \ref{pds}). Therefore, we used an additional Lorentzian to fit the low-frequency noise component whenever required.\\
We have studied the QPO fractional rms amplitude in the 0.5--10.0 keV energy band as a function of QPO frequency (left panel of Figure \ref{qpo_evolution}) and confirmed that the QPO we have identified as fundamental followed a similar relation to the one found for GRS 1915+105 \citep{zh20}. The type-C QPO appears in the LHS and HIMS as a narrow peak with high rms amplitude in the PDS. The properties of the observed broadband noise and the QPO justify the identification of the QPO as type-C \citep{ca04}. We fitted the PDS for three different energy bands (0.5--2.0 KeV, 2.0--4.0 keV, 4.0--10.0 keV) when the type-C QPO was at 1.8 Hz,  4.5 Hz, and 7.0 Hz. We show the fitted PDS and their respective frequency lag spectra in Figure \ref{pds}. The lag and rms values at the QPO frequency are given in Appendix Table \ref{table_appendix}.
When the QPO frequency is higher than 7.0 Hz, the QPO fractional rms amplitude decreases, and the harmonic component becomes insignificant.\\

\begin{figure*}
\centering\includegraphics[width=0.45\textwidth,height=0.25\textheight]{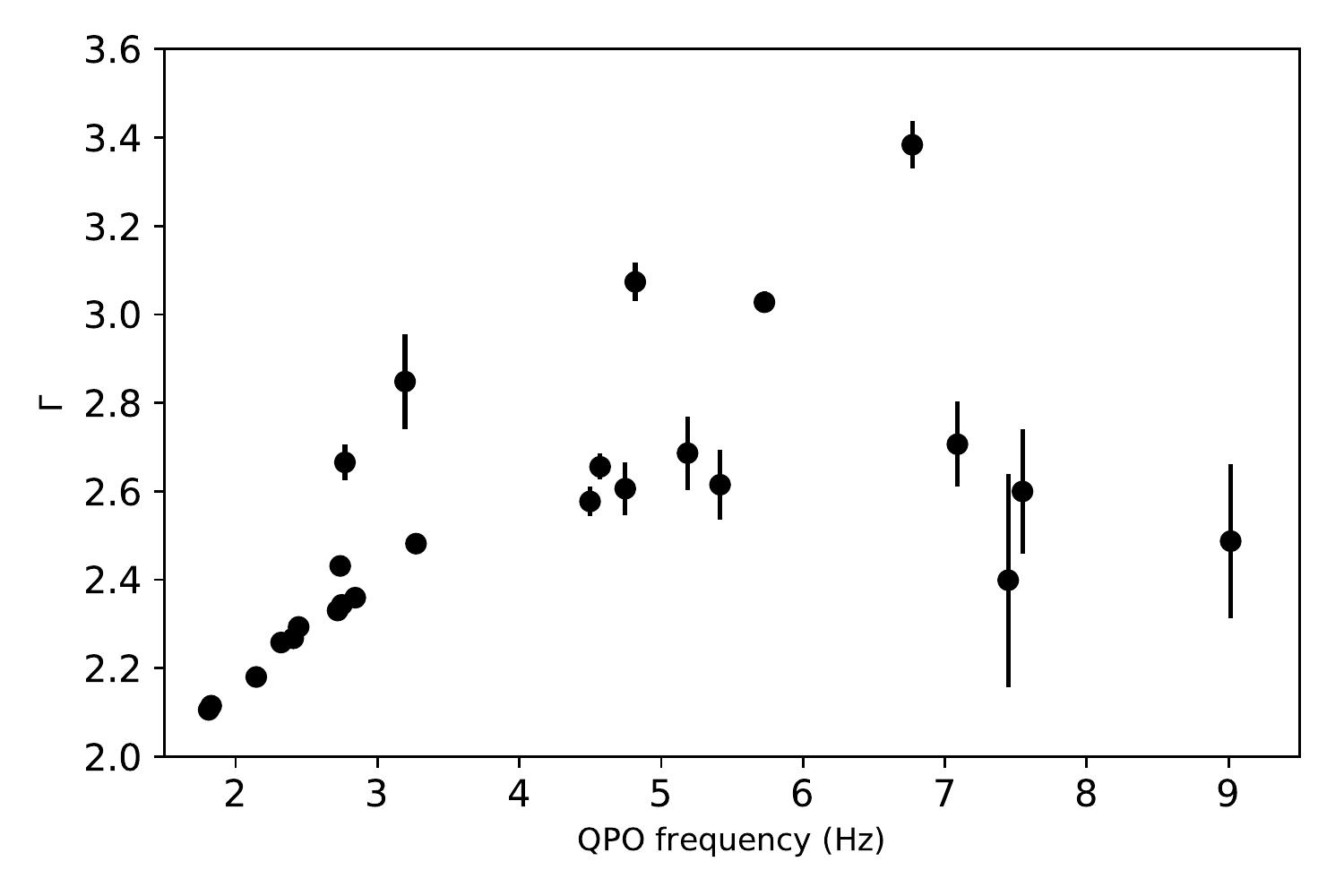}
\centering\includegraphics[width=0.45\textwidth,height=0.25\textheight]{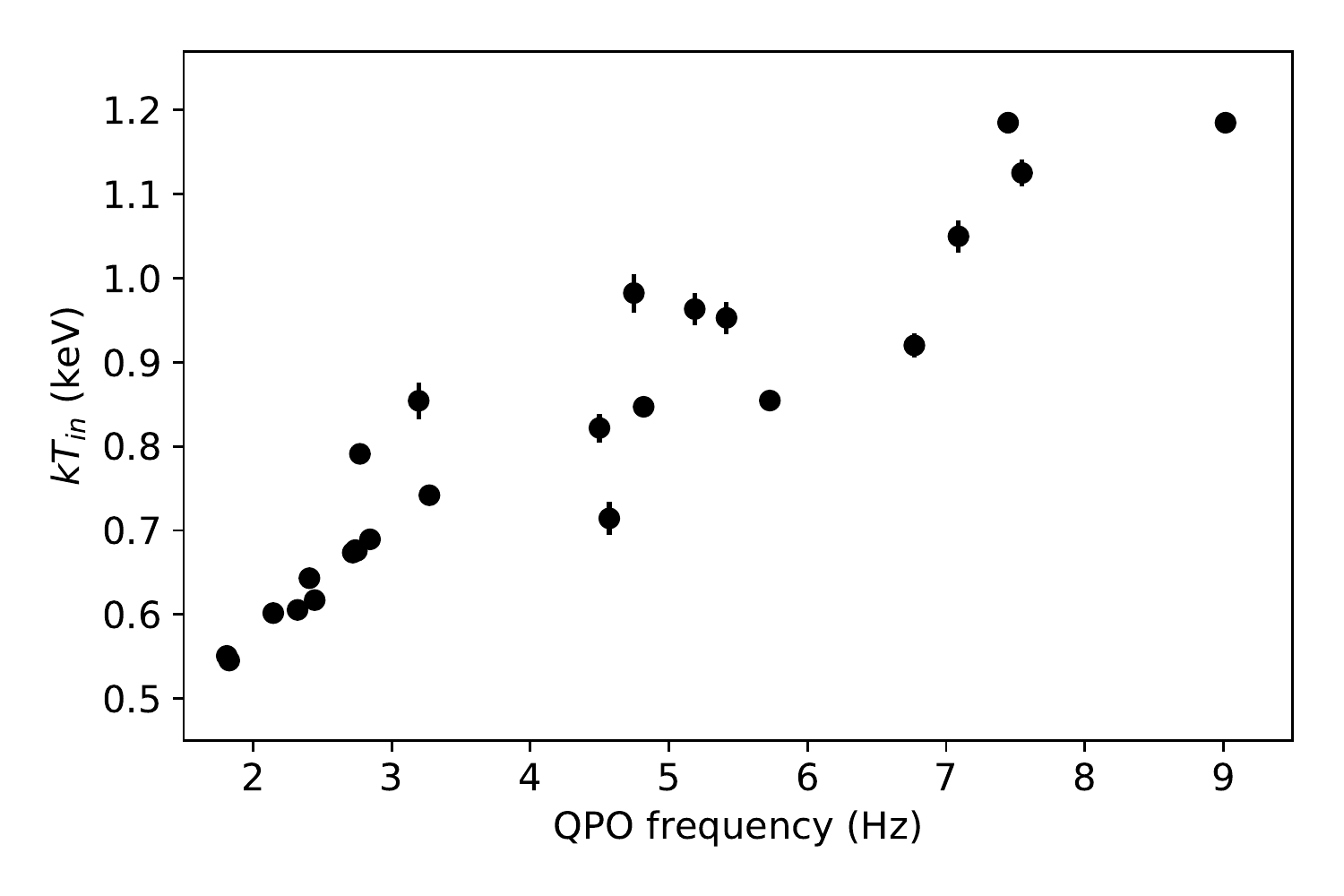}
\caption{The dependence of $\Gamma$ (left panel) and $kT_{in}$ (right panel) upon QPO frequency in MAXI~J1535$-$571. The values of $\Gamma$ and $kT_{in}$ are obtained from the fits to the time-averaged spectra, the rms and phase-lag spectra of the QPO.}
\label{spectral_par}
\end{figure*}

The evolution of the QPO centroid frequency is shown in the right panel of Figure \ref{qpo_evolution}. The QPO frequency first decreases from 2.7 to 1.8 Hz and then increases to its maximum value of 9.0 Hz. After that, the QPO frequency varies in the $4.5-7.5$ Hz range. The QPO frequency and fractional rms amplitude in the $0.5-10.0$ keV band for each observation are given in Table \ref{table1}. We have plotted $\Gamma$ and $kT_{in}$ as a  function of QPO frequency as shown in Figure \ref{spectral_par}. We found that both $\Gamma$ and $kT_{in}$ increase with QPO frequency. 

To extract the rms spectrum, we fit the PDS in 10 energy bands, fixing the QPO centroid frequency and FWHM to the best-fitting values in the 2.0--10.0 keV PDS. The rms and phase lag spectra when the QPO frequency was 1.8 Hz, 4.5 Hz, and 7.0 Hz are shown in the top and bottom panels of Appendix Figure \ref{appendix_fig}. While the fractional rms amplitude of the QPO increases with photon energy for all QPO frequencies, the rms spectrum steepens as the QPO frequency increases from 1.8 Hz to 7.0 Hz (see upper panels in Appendix Figure \ref{appendix_fig}). The change of the slope of the rms spectrum of the QPO is driven by a factor $\sim 3$ drop of the rms amplitude at the lowest energies when the QPO is at low frequencies. In contrast, the rms amplitude at the highest energies remains more or less constant as the QPO frequency changes by a factor of $\sim 4$. Although, in general, the low-energy photons at the QPO frequency lag behind the high-energy photons for all QPO frequencies, the lag spectrum of the QPO changes with QPO frequency. When the QPO frequency is between 1.8 Hz and 2.4 Hz, the lag spectrum shows a minimum at $\sim 4$ keV, with the photons at low and high energies lagging the 4--5 keV photons by $0.1-0.3$ rad. As the QPO frequency increases, the minimum of the lag spectrum of the QPO moves to higher energies, with the minimum reaching $\sim 9-10$ keV at the highest QPO frequency, and the low-energy photons lag the high-energy ones by up to $\sim 0.8$ rad. The rms and phase-lag spectra of the QPO in MAXI ~J1535 in these observations with NICER are consistent with the pattern observed for the type-C QPO by \citet{ra19} in GRS 1915+105 and \citet{ga22} in MAXI ~J1535 with AstroSat, over the common energy range of both instruments. 
\begin{figure}
\centering\includegraphics[scale=0.43,angle=-0]{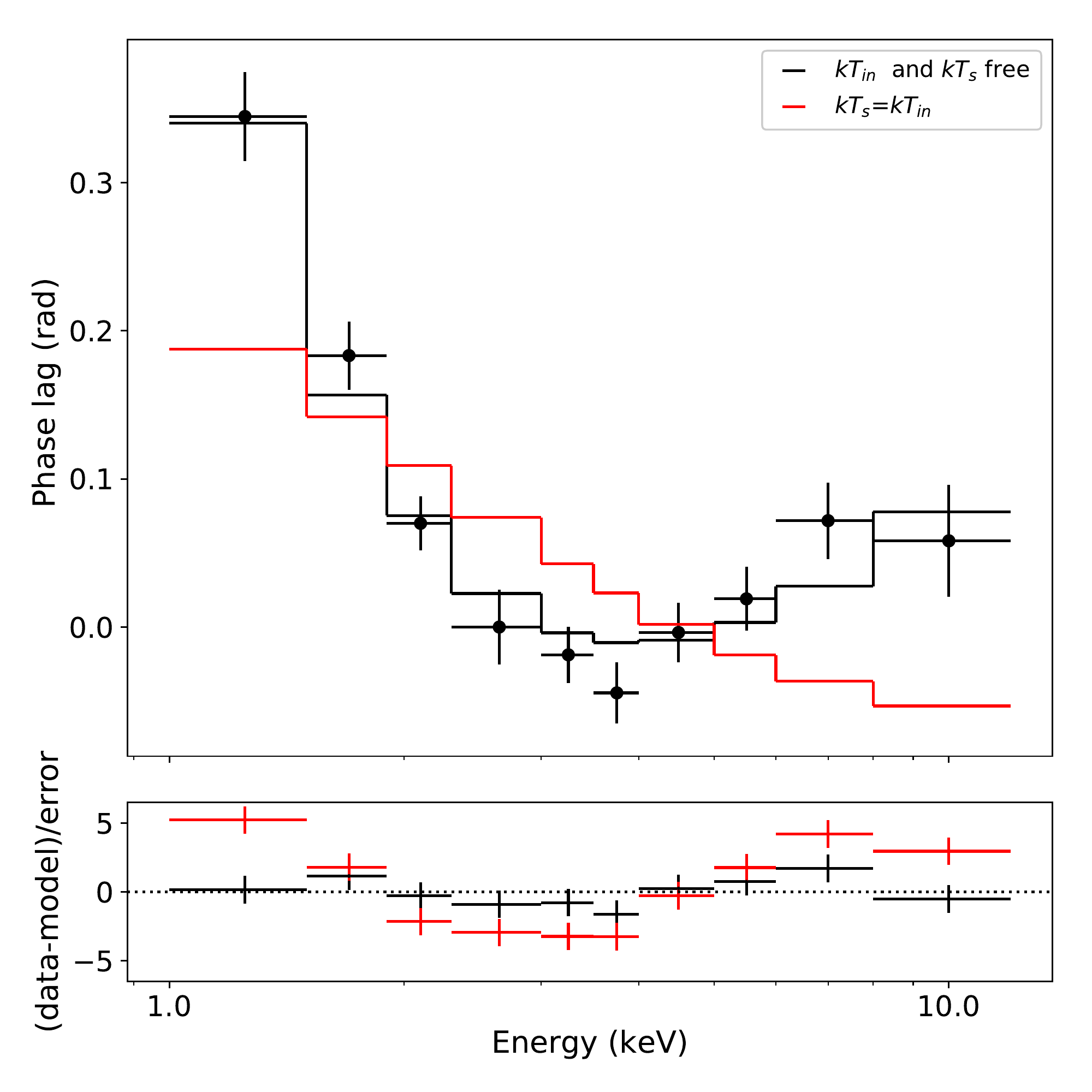}
\caption{The phase-lag spectra of the QPO of MAXI J1535$-$571 fitted with the {\sc vkompthdk} model keeping $kT_{in}$ and $kT_{s}$ tied to each other (red), and free (black). The bottom panel shows the respective residuals of the fits. The data corresponds to obs ID 1050360105 with QPO frequency$\sim$1.8 Hz}
\label{kTin_KTs}
\end{figure} 
\begin{figure*}
\centering\includegraphics[scale=0.65,angle=0]{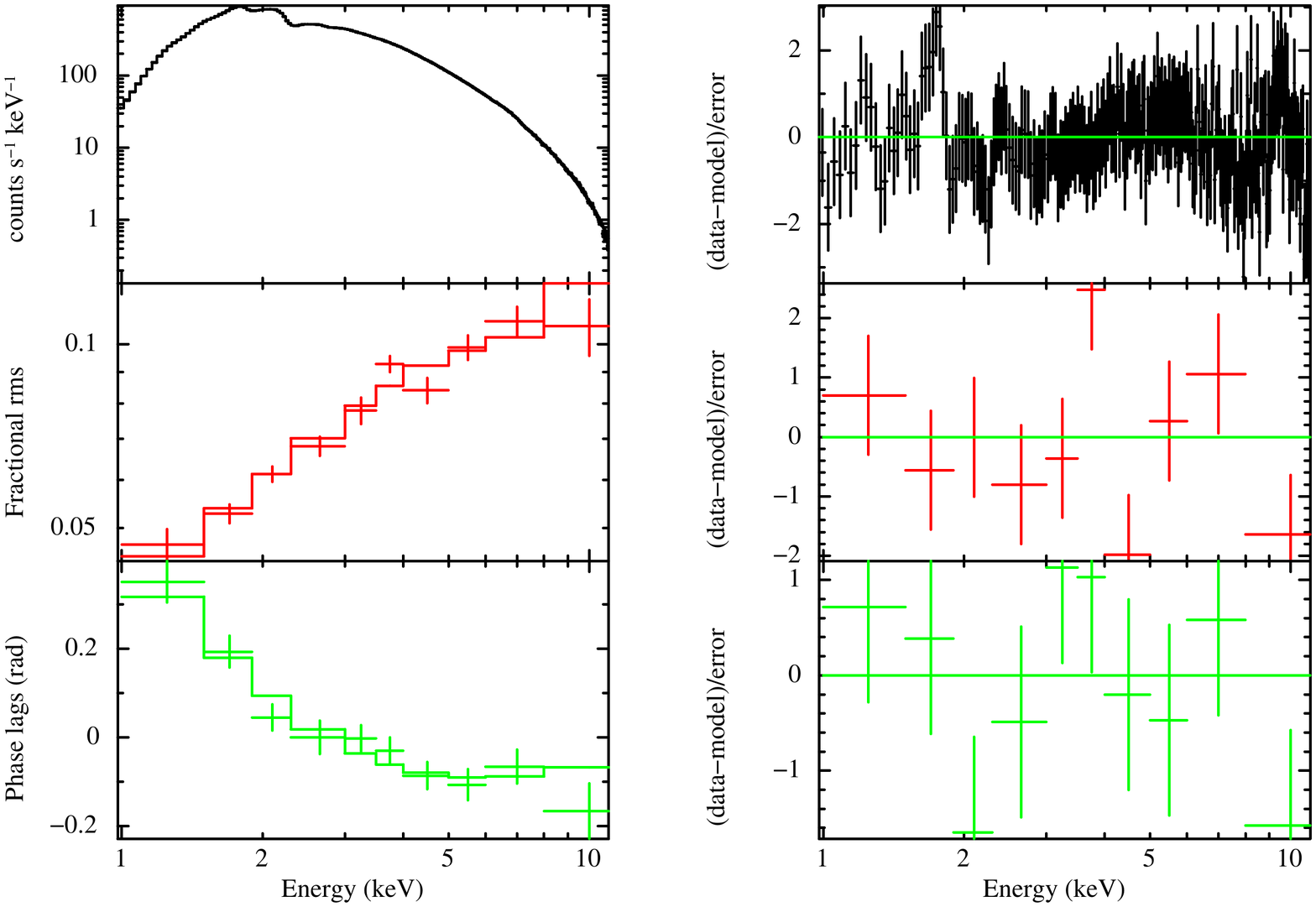}
\caption{Fits of the  {\sc{vkompthdk}} model to the data of MAXI J1535—571. From top to bottom, the left panel shows the time-averaged spectrum of the source fitted with the model  {\sc tbabs*(diskbb+gauss+nthcomp)}, the rms spectrum of the QPO fitted with the model {\sc vkompthdk*dilution}, and the phase-lag spectrum of the QPO fitted with the model {\sc vkompthdk} when the QPO frequency was at $\sim$1.8 Hz. The right panels show the respective residuals of the best-fitting model to the data. The 2.0--3.0 keV band is the reference band for the phase lag spectra.}
\label{rms_lag}
\end{figure*} 
\begin{figure*}
\centering\includegraphics[width=0.49\textwidth,height=0.26\textheight]{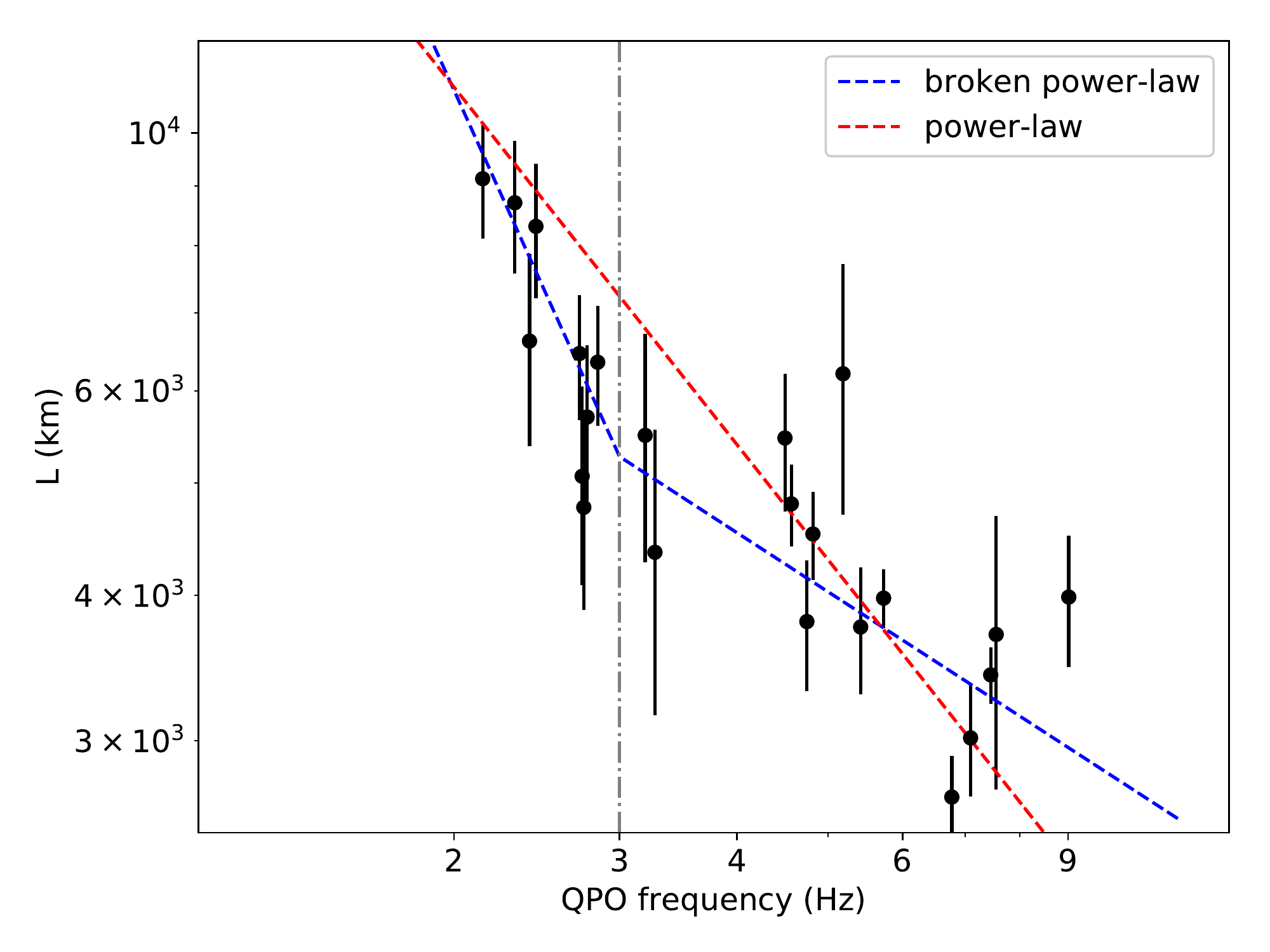}
\centering\includegraphics[width=0.49\textwidth,height=0.26\textheight]{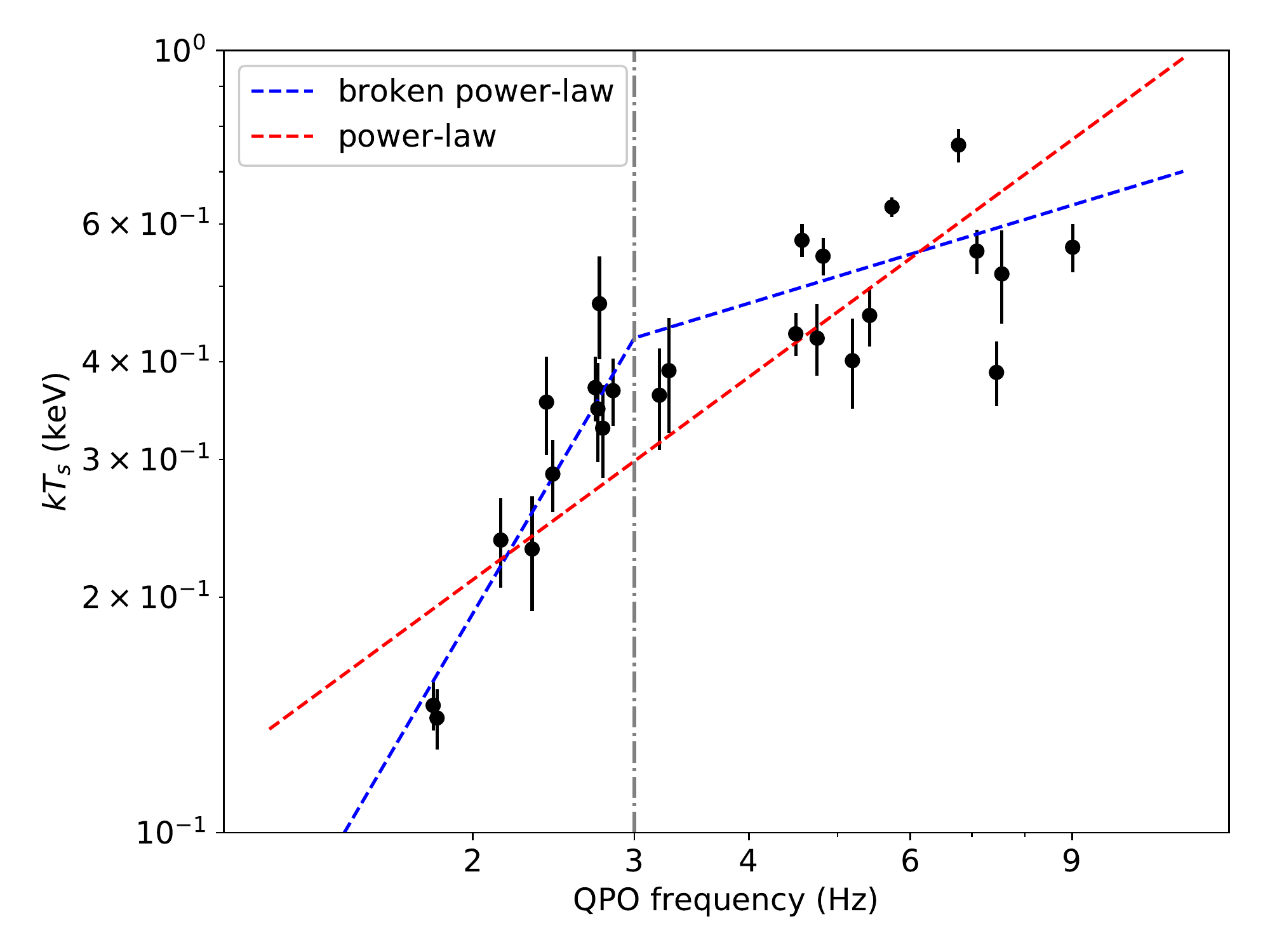}
\centering\includegraphics[width=0.49\textwidth,height=0.26\textheight]{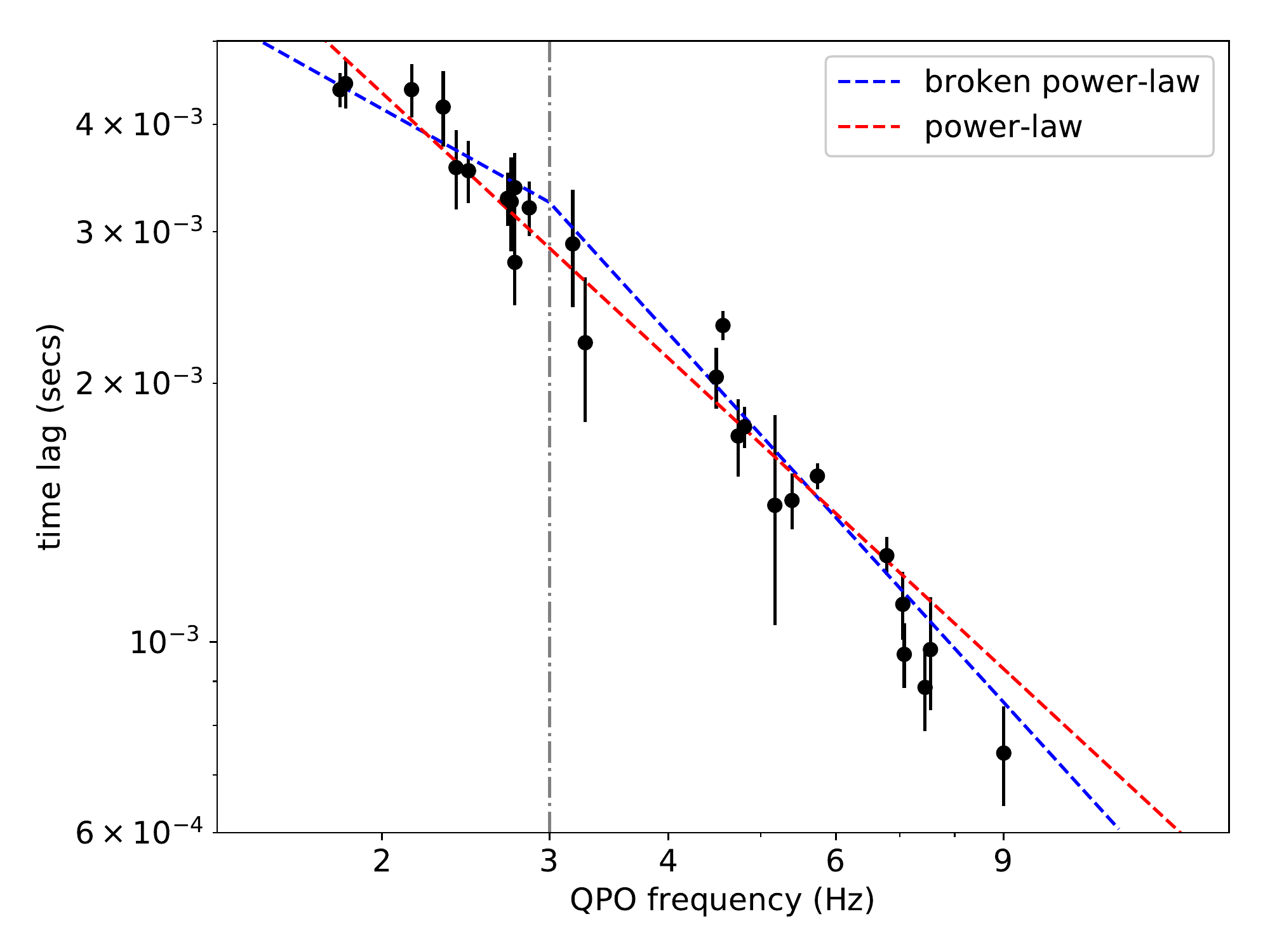}
\centering\includegraphics[width=0.49\textwidth,height=0.26\textheight]{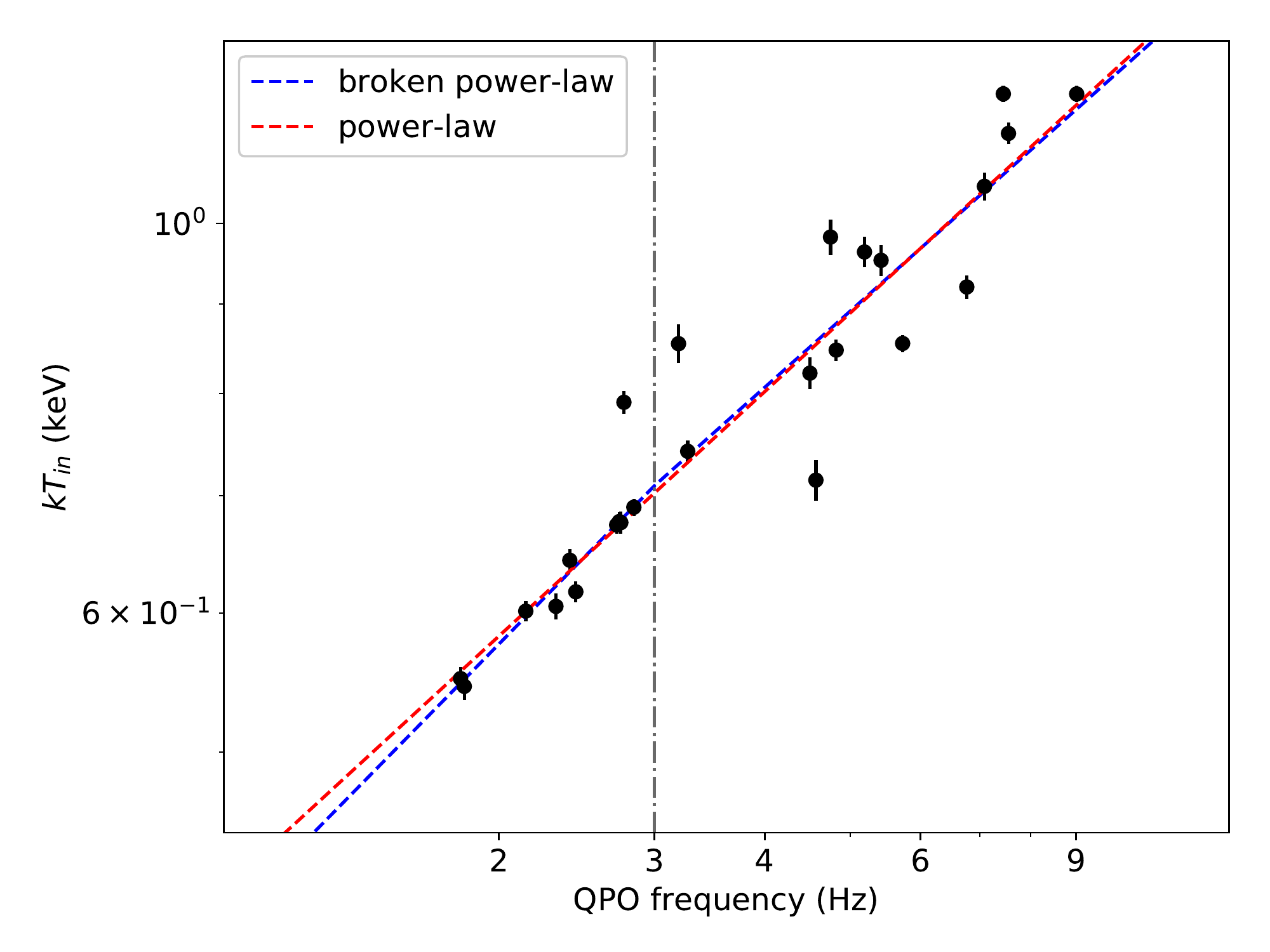}
\caption{Dependence of $L$, $kT_{s}$, time lags of the QPO and $kT_{in}$ upon QPO frequency in MAXI~J1535 $-$571. The red and blue dotted lines show the best-fitting power law and a broken power-law to the data. The best-fitting parameters for each relation are given in Table \ref{table3}. The time lags are between photons in the 1.0--12.0 keV and 2.0--6.0 keV bands at the QPO frequency. The vertical dotted dashed line represents the best-fitting break frequency, $\nu_{c}=$ 3.0 Hz.}
\label{qpo_para}
\end{figure*}
 \subsection{One component time-dependent Comptonization model}
 To understand the changes observed in the rms and lag spectra of the QPO (see Section 3.2), we fitted the rms and lag spectra of the QPO at each QPO frequency with the  {\sc{vkompthdk}} model.
 During the fits we linked $kT_e$ and  $\Gamma$ of {\sc{nthcomp}} to $kT_e$ and  $\Gamma$ of  {\sc{vkompthdk}}.
 We first linked $kT_s$ of {\sc{vkompthdk}} to $kT_{in}$ of {\sc{diskbb}}, and we found large residuals in the fits of the phase-lag spectra (Figure \ref{kTin_KTs}) because {\sc{vkompthdk}} fails to reproduce the minimum of the lags. We subsequently let $kT_{in}$ and $kT_s$ vary independently, and the fits improve significantly (Figure \ref{kTin_KTs}). The simultaneous fitted time-averaged spectra, rms spectra and lag spectra when the QPO frequency was $\sim$1.8 Hz and the residuals of the best-fitting model are shown in Figure \ref{rms_lag} (The peak in the residuals of the time-averaged spectra at 1.84 keV corresponds to the absorption edge features of silicon.). We show a similar plot for the QPO frequencies 4.5 Hz and 7.0 Hz (for which we show a PDS in Figure \ref{pds}) in the Appendix Figures \ref{appendix_fig2} and \ref{appendix_fig3}.
 We discuss the implication of letting $kT_{in}$ and $kT_{s}$ free in Section 4.3.
 The best-fitting parameters and $\chi^{2}$ of the fits are given in Table \ref{table2}.\\
We plotted the model parameters as a function of QPO frequency in Figure \ref{qpo_para}. The size of the corona decreases from $\sim 10^4$ km (which corresponds to 670 R$_g$ for a 10 M$\odot$ black hole ) to $\sim 3\times10^{3}$ km (201 R$_g$) while the temperature of the seed photon source, $kT_s$, increases from $\sim$ 0.1 keV to $\sim$ 0.4 keV as the QPO frequency increases from 1.8 Hz to $\sim$3.0 Hz.  At QPO frequencies $\ge$3.0 Hz, the size of the corona and the temperature of the seed photon source remain more or less constant at respectively $\sim 3-6 \times 10^{3}$ km and 0.5 keV. The error bars on $\eta$ are large, and it is hard to follow any trend if present, although $\eta$ appears to decrease from $\sim$0.8 to $\sim$0.6 as the QPO frequency increases as shown in Appendix Figure \ref{appendix_fig_eta_qpo}. The best-fitting values of $\eta$ imply that $\eta_{int}$ is in the range of 10$-$25$\%$. Comparing the trends in Figures \ref{spectral_par} and \ref{qpo_para}, it is apparent that there is a sudden change of the properties of the source when the QPO frequency is below and above $\sim 3.0$ Hz. The change of behaviour of all the quantities appears to occur at the same QPO frequency, which we call critical frequency, $\nu_{c}$. \\
To estimate the critical frequency, we assume that the break in the relation of the disc and corona model parameters, and time lags as a function of QPO frequency, happens at the same QPO frequency, i.e., $\nu_{c}$. In Figure \ref{qpo_para} we show fits with a power-law (red) and broken power-law (blue) to the relation of $L$, $kT_{s}$, time lag, $kT_{in}$ with QPO frequency. The parameters of the broken power law are the power-law indices $\alpha_1$ and $\alpha_2$ below and above the break frequency $\nu_c$ and a normalisation parameter. We have calculated the F-test probability for the fits with a power law and a broken power-law and found that the probability ranges from ($0.2 - 1)\times 10^{-4}$, which indicates that a broken power-law in general fits the data better than a power law. (To account for the dispersion of the data points around the model was larger than the statistical errors, we have added a systematic of 6$\%$.) The break for each individual fit is in the range 2.7--2.8 Hz, and the break appears to be at the same QPO frequency in all cases. Since there is a hint of a break in the relationship of the time lags and $kT_{in}$ with QPO frequency, we fitted all the four relations ($L$, $kT_{s}$, time lag, $kT_{in}$) together with a broken power law model as shown in Figure \ref{qpo_para}, with the critical frequency tied. We got $\nu_{c}=3.0\pm{0.4}$ Hz. If we let $\nu_{c}$ vary separately for each fit, the $\chi^{2}$ changes from  141.84 (dof=88) to 133.38 (dof=85) with an F-test probability of $\sim$ 0.15. This confirms that the best fit does not improve significantly if we let $\nu_{c}$ free. We conclude that the break is consistent with being at the same frequency in all relations plotted in Figures \ref{spectral_par} and \ref{qpo_para}. The details of the best-fitting parameters are given in Table \ref{table3}.
\begin{table}
 \caption{Broken power-law best-fitting parameters to the relations of $L$, $kT_{s}$, time lags of the QPO and $kT_{in}$ vs. QPO frequency shown in Figure \ref{qpo_para}. The parameters $\alpha1$ and $\alpha2$ are the power-law indices for $\nu_{QPO} \leq \nu_c$ and $\nu_{QPO} > \nu_c$, respectively.}
\begin{center}
\scalebox{0.9}{%
\begin{tabular}{ccccc}
\hline  \hline
Parameter & $\alpha1$ & $\alpha2$ & bknpower norm \\ \hline
L   (km)     &  $1.8\pm{0.4}$  & $0.5\pm{0.2}$  & $(3.8\pm{1.3})\times 10^4$\\
$kT_{s}$ (keV) & -$2.2\pm{0.5}$ & -$0.3\pm{0.2}$ & $0.04\pm{0.01}$\\
$kT_{in}$ (keV) &   -$0.6\pm0.2$  &  -$0.4\pm0.1$  & $0.7\pm0.1$ \\
time lag (m sec)&  $0.6\pm0.4$ & $1.2\pm0.2$  & $0.007\pm0.002$\\
\hline
\end{tabular}}
\end{center}
\footnotesize{Note: The best-fitting parameters values shown above are for the joint fits of all the parameter vs. QPO frequency plot with $\nu_c$ tied.}
\label{table3}
\end{table}

\section{Discussion}
We have analysed NICER observations of MAXI~J1535$-$571 during the initial phase of the outburst in September and October 2017. The rms and lag spectrum of the type-C QPO, the spectral parameters deduced from fits to the time-averaged energy spectra of the source (the temperature of the accretion disc, $kT_{in}$), and the parameters from fits to the rms and lag spectra of the QPO (the size of the corona, $L$, the temperature of the source that provides the seed photons that inverse-Compton scatter in the corona, $kT_s$, all change in a similar manner as the frequency of the type-C QPO increases from 1.8 Hz to 9 Hz. While some of these quantities increase ($kT_{in}$, $kT_s$, phase lags) and others decrease (rms amplitude of the QPO, $L$ ) with increasing QPO frequency, we find that all these quantities show a significant break in the relation at a QPO frequency $\nu_c \sim 3.0$ Hz.\\

At low QPO frequencies, the lag spectrum of the type-C QPO in MAXI~J1535 increases at low and high energies and is minimum at $\sim 4$ keV. This is similar to what is observed for the type-B QPO in the black hole candidate MAXI~J1348$-$630 (\citealt{be20}, \citealt{ga21}). In the case of MAXI~J1348$-$630, \citet{be20} proposed that the fact that photons at energies below $\sim 3$ keV lag behind photons at $\sim 3$ keV is due to down scattering of the photons emitted by the disc in the corona, that they assume is the jet. To reach these conclusions, instead of a black body-like seed spectrum, \citet{be20} assumed a simplified seed-source spectrum that is flat between 2 and 3 keV and does not emit at other energies. Such a spectrum, however, neglects the dilution of the lags caused by black body photons emitted below 2 keV that escape without being up-/down-scattered in the corona. If one considers a more realistic (a black body or a disc) seed spectrum of equivalent temperature, the lags turn out to be flat below $\sim 2-3$ keV, different from what is observed \citep{ky21}. On the other hand, using the model of \citet{ka20}, \citet{ga21} showed that the shape of the lag spectrum (and the rms spectrum as well) of MAXI~J1348$-$630 can be explained by corona photons that impinge back onto the accretion disc and emerge later and at energies below those of the photons that were up-scattered in the corona. This feedback loop between the corona and the disc is the reason for the positive lags between the photons with energies below $\sim 2-3$ keV and those with energies of $\sim 2-3$ keV. At the same time, inverse Compton scattering in the corona explains that photons with energies above $\sim 2-3$ keV lag behind the $2-3$ keV photons. Our fits to the rms and lag spectra of the QPO in MAXI~J1535 here show the same. \\

\subsection{Connection of critical frequency with radio jet quenching}
Using {\sc{AstroSat}}, and {\sc{swift}} observation of the period MJD $58008-58013$ and $58004-58017$, \citet{me18} and \citet{bh19} found a tight correlation between the QPO frequency and the power-law index that models the hard component in the energy spectrum. Using {\sc{nicer}} observation of the period MJD $58008.99-58037.68$, we, on the other hand, found a significant break in the spectral and corona parameters as a function of QPO frequency. The rms and lag spectra of the QPO below and above $\nu_c$ are also significantly different. The break in the relation between the QPO lags and QPO frequency at $\nu_{c} \sim$3.0 Hz in MAXI J1535 is similar to the break found by \citet{zh20} in GRS 1915+105 when the QPO frequency is $\sim$2 Hz, and to the one in GX 339-4 \citep{zh17} at a QPO frequency of $\sim$1.7 Hz.\\
Interestingly, the frequency of the QPO in MAXI~J1535 crosses the value of $3.0$ Hz on September 17 2017 (MJD 58013; see Figure \ref{qpo_evolution} and Table 1). This date coincides with the time at which the radio emission from the jet in this source is quenched \citep{ru19}, which we marked by the shaded area in Figure \ref{qpo_evolution}. Indeed, the radio emission of the jet in MAXI~J1535 quenches in the period MJD $58013.60 - 58014.18$; after that, in the period MJD $58014.18 - 58015.37$ \citep[Table 1][]{ru19} the source makes a transition from the hard intermediate to the soft intermediate state. A similar behaviour has been observed by \citet{me22} for GRS 1915+105, i.e., a low radio emission at or above a QPO frequency of $\sim$2.0 Hz, and increased radio emission below that QPO frequency, the QPO frequency at which \citet{zh20} found that the lags of the QPO change from soft to hard.\\

\subsection{Size of the corona}
From fits to the rms and lag spectra of the QPO with the {\sc{vkompthdk}}, here we find that the size of the corona decreases very rapidly from $\sim 10^4$ km to $\sim 4000-5000$ km when the QPO frequency increases from $\sim 1.8$ Hz to $\sim 3.2$ Hz; from that point on the corona size remains more or less constant or decreases slightly from $\sim 4000-5000$ km down to $\sim 3000$ km as the QPO frequency increases from $\sim 3.2$ Hz up to $\sim 9$ Hz. 
Figure \ref{qpo_evolution} shows that the QPO frequency does not increase monotonically during these observations. In contrast, from Figures \ref{qpo_evolution} and \ref{qpo_para}, it is apparent that the size of the corona first increases from $\sim 2000$ km to $\sim 10^4$ km, and it then decreases back to $\sim 3000$ km (first 10 points in the right panel of Figure \ref{qpo_evolution}). At this time, coincident with the time that the radio emission from the jet is quenched (Russell et al. 2019), the size of the corona continues decreasing but at a lower rate than before. Assuming that MAXI J1535 harbours a 10-solar mass black hole, the maximum and minimum size of the corona are, respectively, $\sim 670$ and $\sim 201$ $R_g$. \\

At low QPO frequency, the trends of the corona size and feedback fraction as a function of QPO frequency reported in this work are similar to those in \citet{zh22}, and both in their work and ours the relation between the size of the corona and the frequency of the QPO shows a break at $\nu_{QPO} \approx 3-4$ Hz. The difference between their and our corona sizes in the common range of QPO frequency comes from the coverage down to lower energies with NICER in our case than in \citet{zh22} with HXMT: The magnitude of the lags of the QPO increases as energy decreases, and the size of the corona in the {\sc{vkompth}} model is driven by the magnitude of the lags. Since we go to lower QPO frequencies than \citet{zh22}, we find that the size of the corona continues increasing as the QPO frequency decreases below $\sim 2$ Hz, where they do not have data. At QPO frequencies above $\sim 4$ Hz \citet{zh22} find an increase of the corona size, whereas here we find that the size continues decreasing with QPO frequency, albeit at a slower rate than below $\sim 3-4$ Hz. We note that \citet{zh22} did not include the effect of dilution of the non-variable components the rms amplitude of the QPO in their model, and that dilution is more important at high QPO frequency, where the contribution of the accretion disc to the total emission increases.\\

Our result is similar to previous findings in other BHXBs (e.g. \citealt{ka19}, \citealt{ka21}). 
In contrast to \citet{ka19} where a change of the vertical size of the corona is proposed to explain the shorter reverberation lags for MAXI J1820+070, \citet{de21} infer a change in the inner accretion disc radius leading to smaller coronal size than reported in this work. Using the JED-SAD model for the same source, \citet{marino21} reported that the size of the jet emitting region, which plays the corona role in their model, of 30-60 $R_g$.  \citet{ax21} showed that the variability of the iron line feature could not be explained using the lamp-post geometry assumed by \citet{ka19} and, instead, a truncated inner hot flow geometry is required. Using a spectral-timing model based on propagating fluctuations and incorporating the reverberation from the variable Comptonisation components, \citet{ka22} further supported a truncated inner hot flow geometry. However, we note that the mass accretion rate propagation fluctuation mechanism used by \citet{ka22} can only explain the hard lags, and a separate mechanism is required to explain to soft lags in MAXI J1820+070 and in the QPO of MAXI J1535$-$571 and other sources.\\

The trend of the size of the corona vs QPO frequency is similar in MAXI~J1535$-$571 and GRS 1915$+$105 (see Figure \ref{qpo_para}, and the supplementary Figure 4 in \citealt{me22} and figure 5 in \citealt{gaf22}). Using a reverberation model for the lags of the broadband noise component in the power spectrum, \citet{wa21} found a corona that is  $\gtrsim$300 $R_g$ in the hard to soft state transition of MAXI J1820+070. Similarly, using polarimetry measurements with PoGO$+$, \citet{ch18} found that the corona in Cyg X-1 is $\gtrsim$100 $R_g$, while they exclude a corona of $\sim$6 $R_g$ obtained from the lamp post model. The sizes reported in this work are consistent with the values published by \citet{ky19}, \citet{ky21}, \citet{re21}, who used Monte Carlo simulations of Comptonization in a jet. The Comptonization model used in this work has some simplifications; for instance, the corona is spherically symmetric with constant temperature and optical depth. This was discussed in \citet{ka21}, and \citet{ga21} and, as explained in \citet{me22}, since the actual geometry of the corona is likely different, the values given by the model should be considered as a characteristic size of the corona rather than the actual radius of a spherical corona (see \citealt{me22,gaf22}).\\

The size of the corona that we infer from our model is larger than the values obtained from fits to the energy spectra of black-hole systems with models that consider reflection off the accretion disc from a corona that is assumed to be a lamppost emitter (e.g., \citealt{vi16}). These spectral fits yield corona sizes of $1-20$ $R_g$ \citep{fb12}. Using the average soft lags over a broad frequency range in the power spectrum and light travel-time arguments, \citet{wa22} found that corona sizes in a dozen black-hole systems in the hard-intermediate state, during the transition from the low-hard to the soft-intermediate state, are comparable, within a factor of a few, to the ones we infer here  (see also \citealt{wa21}). Suppose the assumption that the lags of the broadband noise reflect the light travel time from the corona to the disc is correct. In that case, the corona sizes in \citet{wa22} are, in fact, lower limits for two reasons: (i) \citet{wa22} estimate the corona sizes based on the average time lag over a broad frequency range, whereas the magnitudes of the soft lags are larger than the average over a large range of QPO frequencies (see, for instance, their Fig. 3, panel h). (ii) \citet{wa22} measured the lags between the bands $0.5-1$ and $2-5$ keV. Suppose the lags are minimum at around $\sim 2$ keV and increase both at energies below and above that (see their Fig. 3, panel g). In that case, the magnitude of the time lags between photons at $\sim 2$ and $\sim 0.5$ keV, and hence the light travel distance from the corona to the disc will be larger than what they report. Notice, however, that in \citealt{ka19}, \citealt{wa21} and \citealt{wa22}, the authors estimate the characteristic height of the lamppost corona above the disc.\\

Notice that it is not straightforward to infer sizes from simple light travel-time arguments applied to the time lags of the broadband noise components because: (i) The broadband noise component in the power spectrum of accreting black-hole and neutron-star systems is, in fact, the combination of multiple Lorentzians (e.g., \citealt{ps99}, \citealt{no2000}). Since the properties of these Lorentzians are correlated with each other (e.g., frequency-frequency correlations in \citealt{ps99}) and with the source spectral parameters (e.g., \citealt{vi03,me18,ag06} and references therein), therefore, most likely, these Lorentzians are not just an empirical description of the power spectrum, but each of them represents a relatively well-defined, over a limited frequency range, variability component of the physical properties of the accretion flow. Suppose this decomposition is correct (as suggested by the works cited above). In that case, a more logical and accurate way is to compute the phase lag that results from the combined cross spectra of these Lorentzians in the Fourier real and imaginary space. The phase-lag calculated like that can be different from computed from the average of the cross-spectrum over a broad frequency range (as has been done in many works before, see, e.g. \citealt {no99a,re00,al15,wa22}). If the lags calculated from the Lorentzian decomposition, as suggested above, were due to light travel time, the magnitude of time lags (see, for instance, Fig. \ref{kTin_KTs})  imply large corona sizes. So even combining the lags of the Lorentzians in Fourier space will lead to big corona sizes. ii) It needs to be clarified how to convert time lags into distances using simple light travel-time arguments because the lags depend strongly upon Fourier frequency (e.g., Fig. 3 panel h of \citealt{wa22}). Therefore, there is no single Fourier frequency at which the time lag would represent the correct light travel time that should be used to infer the corona size. (We note that models like RELTRANS, Ingram et al. (2019) calculate the full variability self consistently instead of using simple light travel-time arguments.)\\
Given the typical magnitudes of the lags of the QPO (this paper; \citealt{ka20,ga21,ka21,be22}) or of the broadband noise component (\citealt{wa22}; but see above for the caveats of these measurements) in these systems, any variability model that interprets the observed lags as delays of photons travelling through a medium around a compact object would necessarily yield large corona sizes since time lags of a few hundredths to a few tenths of seconds translate into light travel distances of a few thousand to a few 10,000 km. While propagation of accretion-rate fluctuations \citep{ar06} would yield smaller sizes of the comptonizing region because, in this case, the viscous time scale is at play, propagation of accretion-rate fluctuations only account for hard lags. In contrast, the broadband noise component and the QPOs often show soft lags.\\

Our results are not necessarily inconsistent with the QPO frequency being due to Lense-Thirring Precession (LTP, \citealt{st98}; but see \citealt{ma22}). For instance, \citet{in16} fitted the energy spectra of the BHXRB H1743$-$322 over the cycle of a $\sim$4--5 QPO and concluded that the results are consistent with LTP of an inner hot torus in this source. However, as explained by \citet{in16}, their data could be reproduced equally well if the torus was fixed and it was the disc the one that processed at the Lense–Thirring precession frequency. Their choice of one geometry over the other was based on the fact that the rms spectrum of the QPO is hard, and hence the emission at the QPO frequency could not come from the disc. In the model of \citet{ka20}, the rms spectrum of the QPO is a consequence of inverse-Compton scattering of soft disc photons in the corona \citep[the torus in the scenario of][]{in16}, such that the high rms amplitude values of the QPO at high energies may reflect the variability of the soft disc emission at the Lense–Thirring precession frequency that is inverse-Compton scattered in the corona. This, plus the feedback from the corona to the disc, naturally explain the variability of the iron line discussed by \citet{in16} and the rms spectrum of the QPO. The LTP model and the reverberation model for the lags of the QPO in GRS 1915+105 \citep{na22} also yield a large corona (unless one considers an extra lag due to thermalisation; see \citealt{na22}). Therefore, the LTP model needs to explain how a large corona, which should necessarily extend beyond the disc's inner truncation radius, can precess as a solid body. However, whether the QPO frequency is due to LTP is a matter of debate that needs to be addressed by general relativistic magneto-hydrodynamic (GRMHD) simulations, which is beyond the scope of this paper.\\

\subsection{A Dual Corona}
When we tied the inner-disc temperature of the time-averaged spectra, $kT_{in}$, to the seed-photon temperature of the  {\sc{vkompthdk}} model, $kT_{s}$, our fits could not reproduce the shape of the lag spectrum. Letting these two parameters free yields a significant improvement in the fit statistics (see Section 3.3 and Figure \ref{kTin_KTs}). We speculate that this difference between the seed photon temperature of {\sc{nthcomp}} and {\sc{vkompthdk}} is due to a more complex structure of the comptonizing region than that described by a uniform corona. 
\citet{sr19}, \citet{bh19} $\&$ \citet{ga22} used AstroSat observations of MAXI~J1535 that coincide with the first few days of the NICER observations reported in this work. They modelled the combined {\sc{SXT}} and {\sc{LAXPC}} spectra and reported a lower inner disc temperature ($kT_{in}$=0.20--0.35 keV) than we found in this work. It should be noted that \citet{bh19} and \citet{ga22} modelled the spectra in the 1-30 keV energy range. Also, the source is highly absorbed, and the spectrum drops at low energies, so the reported inner disc temperature may not be accurate. \citet{sh19} used the same AstroSat observation and modelled the broadband spectra in the 0.3-80.0 keV band and reported electron temperatures with {\sc{nthcomp}} in the range 21-63 keV. Using the same AstroSat observation, \citet{sr19} reported an electron temperature of $\sim$21 keV. As the 0.8-10.0 keV spectra of NICER could not constrain the electron temperature, we chose to fix it to the values reported by \citet{sh19} and \citet{sr19}. The electron temperature ($\sim$90--108 keV) reported by \citealt{ga22} is higher than the value ($\sim$21 keV) we have used in this work. It should be noted that in \citet{ga22}, they are fixed the optical depth of the corona, which together with $\Gamma$ gives $kT_e$.\\

 Using a dual-component comptonization model for type-B QPOs,
 \citet{ga21} and \citet{pe22} argued that the comptonizing medium of the BHXB sources, MAXI~J1348$-$630 and GX~339$-$4 consist of two coronas. A relatively small corona of $\sim$300 km, close to the black hole dominates the time-averaged spectra, and a large corona of $\sim$18000 km, possibly the jet, dominates the lag spectrum \citep{pe22}. Their best-fitting results yield a lower seed photon temperature of the large corona compared to the small corona, with the seed photon temperature of the small corona linked to $kT_{bb}$ of {\sc{nthcomp}}. \citet{pe22} proposed that this difference is due to the fact that the seed photons for the small corona come from the inner, hotter parts, of the disc whereas the seed photons for the large corona come from the outer, cooler parts, of the disc. 
 A similar dual-corona geometry could explain
 the difference between $kT_{in}$  of the {\sc{diskbb}} (linked to $kT_{bb}$ of {\sc{nthcomp}}) and $kT_{s}$ of {\sc{vkompthdk}} in our fits.
 Since we find that $kT_{bb}>kT_{s}$, also in MAX J1535$-$571 the small corona would dominate the emission of the time-averaged spectra, whereas the big corona would dominate the lags. We found that the rms spectra do not change much between the two fits ($kT_s$=$kT_{in}$ or $kT_s$ free), so we conclude that the rms amplitude is not affected much by the size of the corona. The fraction of the corona flux that returns to the disc is $\eta_{int}$ 10--25 $\%$ in all the cases. This and the large corona size further indicate that the large corona is the jet.
\section{Summary and Conclusions}
We have analysed all NICER observation of MAXI J1535$-$571 taken on September and October 2017. We fit the energy spectra of the source and the rms and lag spectra of the type-C QPO in this source with the one-component time dependent Comptonization model {\sc{vkompthdk}}. Below we summarize our results:
\begin{itemize}
 \item The size of the corona of MAXI J1535$-$571 decreases from $10^4$ km when the QPO frequency is $\ge$2 Hz to $\sim$3000 km when the QPO frequency is $\sim$9.0 Hz.
 
 \item   The behaviour of all the spectral parameters and the rms and lag spectra of the QPO changes above and below a critical QPO frequency, $\nu_c =$3.0$\pm 0.4$ Hz. Interestingly, the time at which this critical frequency happens coincide with the period when the radio jet emission quenches for this source.
\item Comparing our results with those in previous work, the data are consistent with a dual corona: a small corona lying close to the black hole and a larger one, possibly the jet.
   \end{itemize}

\section*{Acknowledgements}
This research is part of a project proposed for the COSPAR PCB fellowship program. We would like to thank the referee for constructive comments that helped improve this paper. DR would like to thank COSPAR, ISRO and Professor Diego Altamirano for jointly funding the academic visit to the University of Southampton. MM, FG and KK acknowledge support from the research programme Athena with project number 184.034.002, which is (partly) financed by the Dutch Research Council (NWO). FG acknowledges support from PIP 0102 and PIP 0113 (CONICET). FG is a CONICET researcher. This work received financial support from PICT-2017-2865 (ANPCyT). KA acknowledges support from a UGC-UKIERI Phase 3 Thematic Partnership (UGC-UKIERI-2017-18-006; PI: P. Gandhi). TMB acknowledges financial contribution from PRIN INAF 2019 n.15. CB is a fellow of Consejo Interuniversitario Nacional (CIN). 

\section*{Data Availability}
The NICER XTI observations used in this work are available at NICER Archive\footnote{\url{https://heasarc.gsfc.nasa.gov/docs/nicer/nicer_archive.html}}.

\def\aj{AJ} \def\actaa{Acta Astron.}  \def\araa{ARA\&A} \def\apj{ApJ}
\def\apjl{ApJ} \def\apjs{ApJS} \def\ao{Appl.~Opt.}  \def\apss{Ap\&SS}
\def\aap{A\&A} \def\aapr{A\&A~Rev.}  \def\aaps{A\&AS} \def\azh{AZh}
\def\baas{BAAS} \def\bac{Bull. astr. Inst. Czechosl.}
\def\caa{Chinese Astron. Astrophys.}  \def\cjaa{Chinese
  J. Astron. Astrophys.}  \def\icarus{Icarus} \def\jcap{J. Cosmology
  Astropart. Phys.}  \def\jrasc{JRASC} \def\mnras{MNRAS}
\def\memras{MmRAS} \def\na{New A} \def\nar{New A Rev.}
\def\pasa{PASA} \def\pra{Phys.~Rev.~A} \def\prb{Phys.~Rev.~B}
\def\prc{Phys.~Rev.~C} \def\prd{Phys.~Rev.~D} \def\pre{Phys.~Rev.~E}
\def\prl{Phys.~Rev.~Lett.}  \def\pasp{PASP} \def\pasj{PASJ}
\def\qjras{QJRAS} \def\rmxaa{Rev. Mexicana Astron. Astrofis.}
\def\skytel{S\&T} \def\solphys{Sol.~Phys.}  \def\sovast{Soviet~Ast.}
\def\ssr{Space~Sci.~Rev.}  \def\zap{ZAp} \def\nat{Nature}
\def\iaucirc{IAU~Circ.}  \def\aplett{Astrophys.~Lett.}
\def\apspr{Astrophys.~Space~Phys.~Res.}
\def\bain{Bull.~Astron.~Inst.~Netherlands}
\def\fcp{Fund.~Cosmic~Phys.}  \def\gca{Geochim.~Cosmochim.~Acta}
\def\grl{Geophys.~Res.~Lett.}  \def\jcp{J.~Chem.~Phys.}
\def\jgr{J.~Geophys.~Res.}
\def\jqsrt{J.~Quant.~Spec.~Radiat.~Transf.}
\def\memsai{Mem.~Soc.~Astron.~Italiana} \def\nphysa{Nucl.~Phys.~A}
\def\physrep{Phys.~Rep.}  \def\physscr{Phys.~Scr}
\def\planss{Planet.~Space~Sci.}  \def\procspie{Proc.~SPIE}
\let\astap=\aap \let\apjlett=\apjl \let\apjsupp=\apjs \let\applopt=\ao

\bibliographystyle{mnras}
\bibliography{manuscript} 

%








\label{lastpage}
\appendix
\counterwithin{table}{section}
\section{ }

\begin{table*}
 \centering
 \caption{The columns are the observation number, the chi-square of the fit to the steady-state spectrum ($\chi^{2}_{SSS}$), rms spectrum ($\chi^{2}_{rms}$), lag spectrum ($\chi^{2}_{lag}$) with, in each case, the number of channels in each spectrum and the total reduced chi-square of the combined fit with degree of freedom.}
 
\begin{center}
\scalebox{1.0}{%
\hspace{-0.5cm}
\begin{tabular}{ccccc}
\hline  \hline
 Obs no. & $\chi^{2}_{SSS}$ (channel) &	$\chi^{2}_{rms}$ (channel) &	$\chi^{2}_{lag}$ (channel) &	$\chi^{2}_{total}$	(dof)\\ \hline
 1 & $206.9 \;(238)$ & $15.5 \;(10)$ & $9.0 \;(10)$ & $231.4 \;(243)$ \\
2 & $176.5 \;(237)$ & $7.8 \;(10)$ & $7.6 \;(10)$ & $191.9 \;(242)$ \\
3 & $219.5 \;(238)$ & $7.7 \;(10)$ & $13.3 \;(10)$ & $240.5 \;(243)$ \\
4 & $205.6 \;(238)$ & $4.7 \;(10)$ & $9.4 \;(10)$ & $219.8 \;(243)$ \\
5 & $206.8 \;(238)$ & $13.8 \;(10)$ & $21.8 \;(10)$ & $242.3 \;(243)$ \\
6 & $167.9 \;(238)$ & $5.1 \;(10)$ & $4.8 \;(10)$ & $177.9 \;(243)$ \\
7 & $165.9 \;(238)$ & $5.0 \;(10)$ & $2.4 \;(10)$ & $173.2 \;(243)$ \\
8 & $227.7 \;(238)$ & $4.8 \;(10)$ & $2.3 \;(10)$ & $234.8 \;(243)$ \\
9 & $157.1 \;(238)$ & $5.0 \;(10)$ & $7.2 \;(10)$ & $169.3 \;(243)$ \\
10 & $146.1 \;(238)$ & $4.7 \;(10)$ & $4.3 \;(10)$ & $155.1 \;(243)$ \\
11 & $176.4 \;(217)$ & $13.0 \;(10)$ & $2.7 \;(10)$ & $192.2 \;(222)$ \\
12 & $129.3 \;(238)$ & $10.6 \;(10)$ & $12.5 \;(10)$ & $152.4 \;(243)$ \\
13 & $157.3 \;(238)$ & $7.3 \;(10)$ & $11.8 \;(10)$ & $176.3 \;(243)$ \\
14 & $147.0 \;(238)$ & $17.7 \;(10)$ & $3.8 \;(10)$ & $168.4 \;(242)$ \\
15 & $183.9 \;(235)$ & $9.3 \;(10)$ & $2.7 \;(10)$ & $195.8 \;(239)$ \\
16 & $146.9 \;(238)$ & $13.3 \;(7)$ & $4.8 \;(7)$ & $165.0 \;(236)$ \\
17 & $142.4 \;(238)$ & $23.0 \;(10)$ & $11.6 \;(10)$ & $177.0 \;(242)$ \\
18 & $240.5 \;(231)$ & $3.0 \;(7)$ & $0.9 \;(7)$ & $244.4 \;(229)$ \\
19 & $184.0 \;(238)$ & $10.5 \;(10)$ & $9.1 \;(10)$ & $203.6 \;(242)$ \\
20 & $185.3 \;(235)$ & $3.7 \;(10)$ & $15.5 \;(10)$ & $204.5 \;(240)$ \\
21 & $181.6 \;(216)$ & $11.6 \;(7)$ & $5.1 \;(7)$ & $198.2 \;(215)$ \\
22 & $211.3 \;(214)$ & $23.1 \;(10)$ & $23.6 \;(10)$ & $258.0 \;(219)$ \\
23 & $183.8 \;(232)$ & $26.2 \;(10)$ & $13.1 \;(11)$ & $223.1 \;(238)$ \\
24 & $184.1 \;(238)$ & $5.0 \;(10)$ & $2.3 \;(9)$ & $191.4 \;(241)$ \\
25 & $159.6 \;(238)$ & $5.2 \;(10)$ & $10.1 \;(10)$ & $174.9 \;(242)$ \\
\hline
\end{tabular}}
\end{center}
  \small
    Note: Notice that some parameters are linked in the combined fits and therefore we cannot give the number of degrees of freedom for each individual fit. So, channel numbers for individual spectra are given here.

\label{table_appendix0}
\end{table*}
\begin{table*}
 \centering
 \caption{The columns are the observation number, QPO frequency, QPO fractional rms amplitude and time lags at the QPO frequency of MAXI J1535$-$571. Here rms1 and lag1 are in the 0.5--2.0 keV band, rms2 and lag2 are in the 2.0--4.0 keV band, and rms3 and lag3 are in the 4.0--10.0 keV band. The reference band for lags is 0.5--10.0 keV.}
\begin{center}
\scalebox{1.0}{%
\hspace{-0.5cm}
\begin{tabular}{cccccccc}
\hline  
\hline
Obs no. & QPO frequency & QPO fractional  &      lag1        &     QPO fractional  &        lag2           &     QPO fractional       &   lag3\\  &(Hz) & rms1  (\%) & (msec)  &  rms2 (\%) & (msec) &  rms3 
(\%) & (msec)  \\ \hline
1 & $2.74 \pm 0.01$ & $5.2 \pm 0.1$ & $10.2 \pm 1.0$ & $7.3 \pm 0.2$ & $-1.49 \pm 0.38$ & $9.4 \pm 0.3$ & $-6.4 \pm 0.7$ \\
2 & $2.44 \pm 0.01$ & $5.0 \pm 0.2$ & $12.5 \pm 0.9$ & $6.7 \pm 0.2$ & $-2.22 \pm 0.41$ & $8.7 \pm 0.3$ & $-7.1 \pm 0.7$ \\
3 & $2.32 \pm 0.01$ & $5.5 \pm 0.2$ & $12.7 \pm 1.2$ & $6.8 \pm 0.3$ & $-3.20 \pm 0.54$ & $8.8 \pm 0.4$ & $-6.0 \pm 1.1$ \\
4 & $1.83 \pm 0.01$ & $5.8 \pm 0.1$ & $12.5 \pm 0.8$ & $7.4 \pm 0.2$ & $-4.63 \pm 0.38$ & $8.7 \pm 0.3$ & $-2.7 \pm 0.7$ \\
5 & $1.81 \pm 0.00$ & $5.8 \pm 0.1$ & $12.1 \pm 0.5$ & $7.4 \pm 0.1$ & $-4.20 \pm 0.22$ & $9.2 \pm 0.1$ & $-3.2 \pm 0.4$ \\
6 & $2.15 \pm 0.01$ & $5.6 \pm 0.2$ & $14.0 \pm 0.9$ & $7.1 \pm 0.2$ & $-3.24 \pm 0.39$ & $8.6 \pm 0.3$ & $-7.1 \pm 0.7$ \\
7 & $2.41 \pm 0.01$ & $5.8 \pm 0.2$ & $13.3 \pm 1.2$ & $7.7 \pm 0.3$ & $-1.59 \pm 0.47$ & $9.8 \pm 0.4$ & $-9.4 \pm 0.9$ \\
8 & $2.77 \pm 0.01$ & $5.5 \pm 0.2$ & $12.6 \pm 1.1$ & $7.6 \pm 0.2$ & $-2.05 \pm 0.42$ & $9.5 \pm 0.4$ & $-6.9 \pm 0.9$ \\
9 & $2.75 \pm 0.02$ & $5.3 \pm 0.2$ & $12.3 \pm 1.3$ & $7.2 \pm 0.2$ & $-1.35 \pm 0.57$ & $10.0 \pm 0.4$ & $-8.4 \pm 1.1$ \\
10 & $3.27 \pm 0.02$ & $4.9 \pm 0.2$ & $9.1 \pm 1.5$ & $7.1 \pm 0.3$ & $-1.44 \pm 0.54$ & $10.6 \pm 0.4$ & $-5.5 \pm 1.0$ \\
11 & $3.19 \pm 0.03$ & $5.3 \pm 0.3$ & $12.6 \pm 1.7$ & $7.0 \pm 0.3$ & $-1.42 \pm 0.65$ & $10.5 \pm 0.5$ & $-7.1 \pm 1.1$ \\
12 & $2.72 \pm 0.01$ & $4.7 \pm 0.2$ & $13.7 \pm 0.9$ & $6.9 \pm 0.2$ & $-1.79 \pm 0.33$ & $9.3 \pm 0.3$ & $-8.1 \pm 0.6$ \\
13 & $2.84 \pm 0.01$ & $5.4 \pm 0.2$ & $13.1 \pm 0.9$ & $7.6 \pm 0.2$ & $-2.10 \pm 0.32$ & $10.4 \pm 0.3$ & $-6.7 \pm 0.6$ \\
14 & $4.75 \pm 0.01$ & $3.2 \pm 0.3$ & $9.3 \pm 0.8$ & $5.6 \pm 0.1$ & $0.23 \pm 0.24$ & $9.7 \pm 0.2$ & $-6.2 \pm 0.4$ \\
15 & $9.01 \pm 0.04$ & $--$ & $4.4 \pm 0.4$ & $1.5 \pm 0.1$ & $0.07 \pm 0.15$ & $3.7 \pm 0.1$ & $-3.2 \pm 0.2$ \\
16 & $7.54 \pm 0.05$ & $1.4 \pm 0.4$ & $6.4 \pm 0.6$ & $2.2 \pm 0.3$ & $0.50 \pm 0.26$ & $6.0 \pm 0.2$ & $-4.7 \pm 0.3$ \\
17 & $7.54 \pm 0.06$ & $1.3 \pm 0.2$ & $5.3 \pm 0.5$ & $2.8 \pm 0.1$ & $0.20 \pm 0.14$ & $5.9 \pm 0.2$ & $-3.9 \pm 0.2$ \\
18 & $7.09 \pm 0.03$ & $1.1 \pm 0.1$ & $4.8 \pm 0.4$ & $2.2 \pm 0.1$ & $0.01 \pm 0.12$ & $5.3 \pm 0.1$ & $-3.6 \pm 0.2$ \\
19 & $5.42 \pm 0.01$ & $2.7 \pm 0.1$ & $7.9 \pm 0.5$ & $4.6 \pm 0.1$ & $-0.21 \pm 0.17$ & $9.3 \pm 0.2$ & $-4.6 \pm 0.2$ \\
20 & $5.73 \pm 0.01$ & $2.6 \pm 0.1$ & $8.3 \pm 0.2$ & $4.4 \pm 0.1$ & $-0.40 \pm 0.08$ & $9.1 \pm 0.1$ & $-4.3 \pm 0.1$ \\
21 & $6.77 \pm 0.02$ & $1.9 \pm 0.1$ & $6.4 \pm 0.3$ & $3.3 \pm 0.1$ & $-0.24 \pm 0.10$ & $7.6 \pm 0.1$ & $-3.7 \pm 0.1$ \\
22 & $4.57 \pm 0.01$ & $2.8 \pm 0.1$ & $10.8 \pm 0.4$ & $4.6 \pm 0.1$ & $-0.91 \pm 0.13$ & $8.2 \pm 0.2$ & $-5.6 \pm 0.2$ \\
23 & $4.82 \pm 0.01$ & $2.0 \pm 0.1$ & $9.5 \pm 0.5$ & $4.0 \pm 0.0$ & $-0.39 \pm 0.13$ & $6.3 \pm 0.1$ & $-5.3 \pm 0.2$ \\
24 & $5.19 \pm 0.03$ & $2.0 \pm 0.2$ & $7.7 \pm 1.7$ & $2.9 \pm 0.2$ & $-0.23 \pm 0.51$ & $7.2 \pm 0.3$ & $-4.6 \pm 0.8$ \\
25 & $4.50 \pm 0.01$ & $3.1 \pm 0.1$ & $9.1 \pm 0.6$ & $5.2 \pm 0.1$ & $-0.69 \pm 0.22$ & $9.2 \pm 0.2$ & $-5.1 \pm 0.4$ \\

\hline
\end{tabular}}
\end{center}
\label{table_appendix}
\end{table*}

\begin{figure*}
\centering\includegraphics[scale=0.43,angle=0]{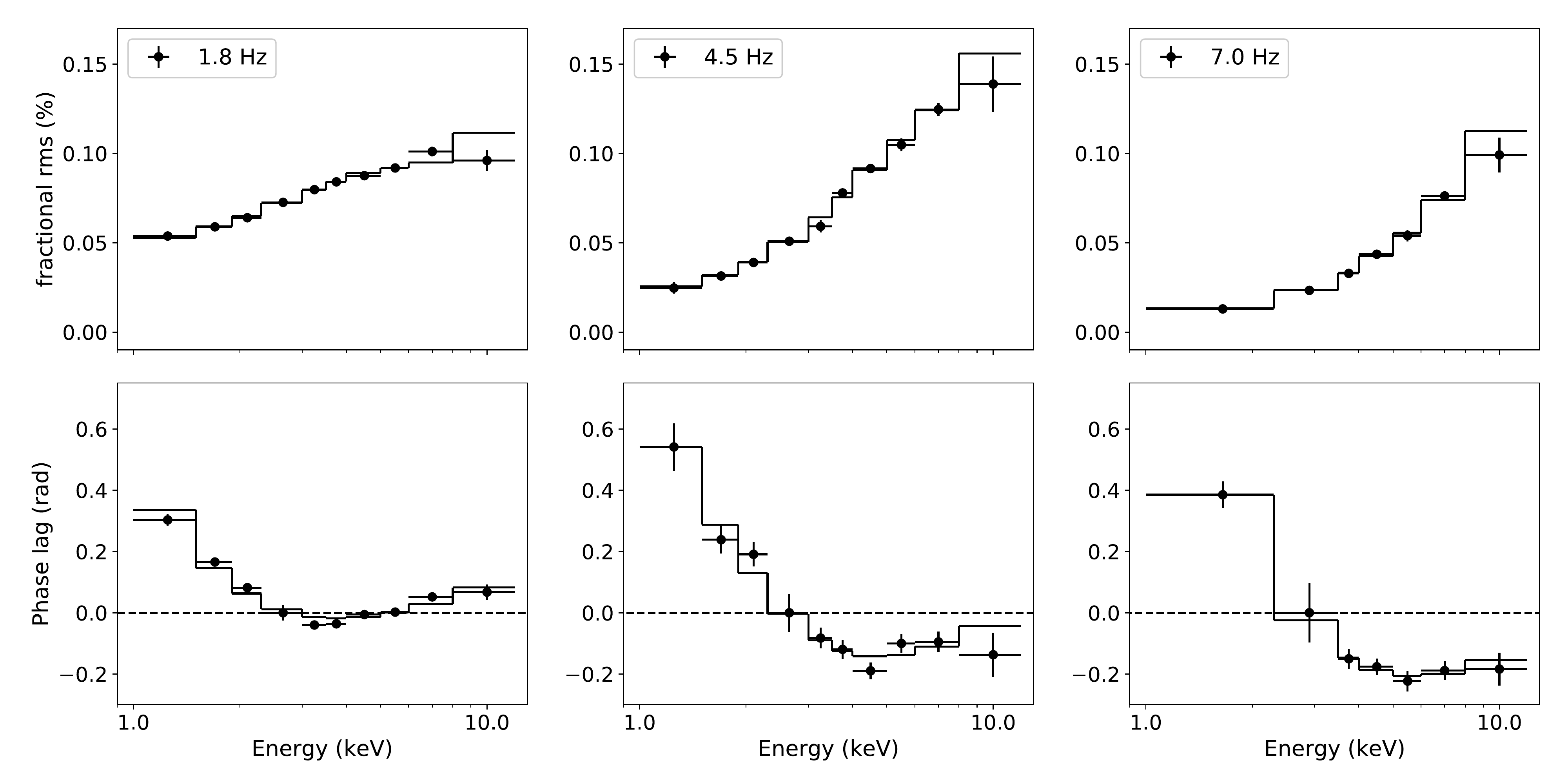}
\caption{The top and bottom panels show respectively the fractional rms and phase-lag spectra of the type-C QPO in MAXI~J1535$-$571 fitted with {\sc{vkompthdk}} model. The 2.0--3.0 keV band is the reference band for the phase lag spectra.}
\label{appendix_fig}
\end{figure*}

\begin{figure*}
\centering\includegraphics[scale=0.65,angle=0]{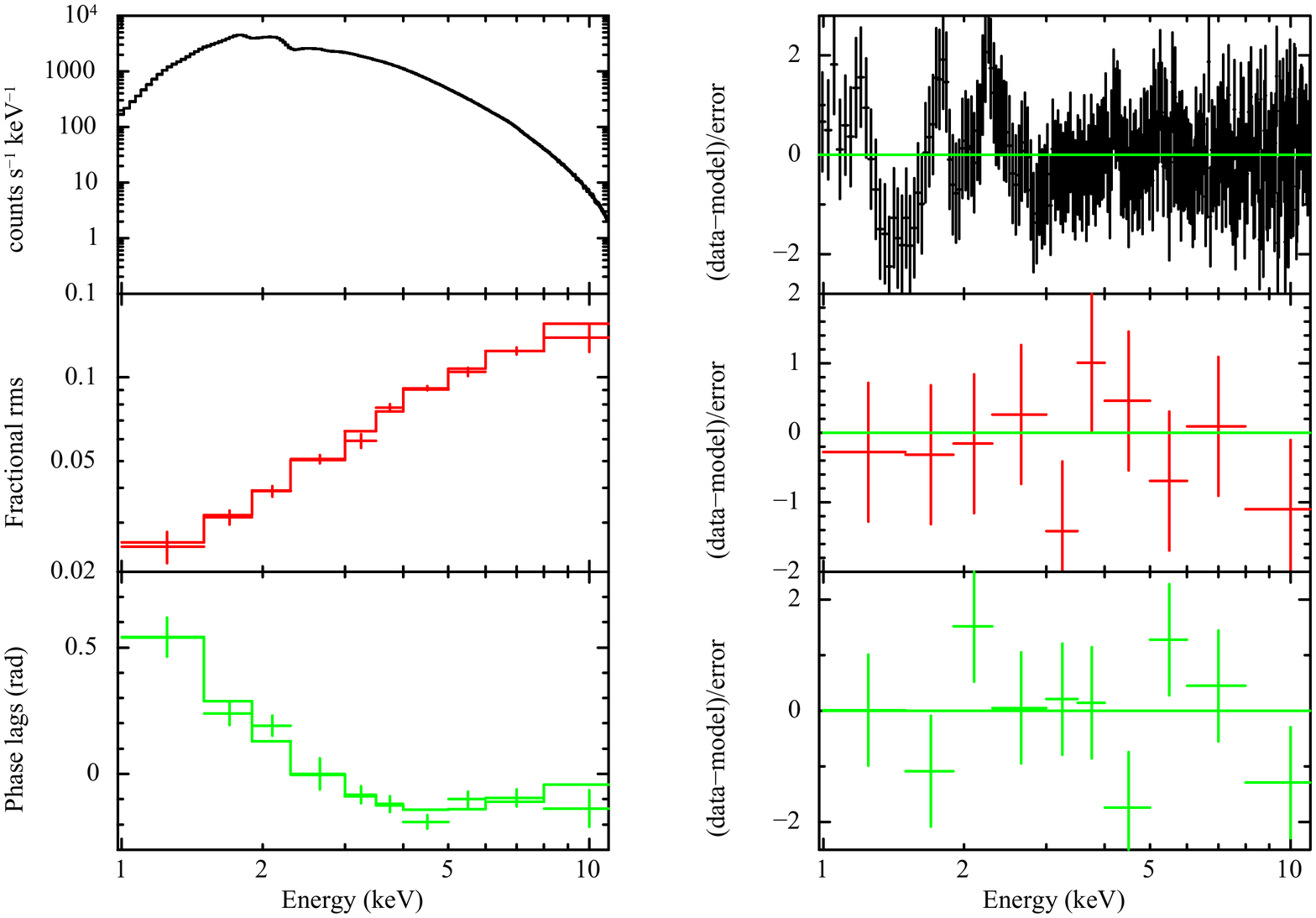}
\caption{The same plot as shown in Figure \ref{rms_lag} at $\sim$4.5 Hz QPO frequency in MAXI~J1535$-$571.}
\label{appendix_fig2}
\end{figure*} 

\begin{figure*}
\centering\includegraphics[scale=0.65,angle=0]{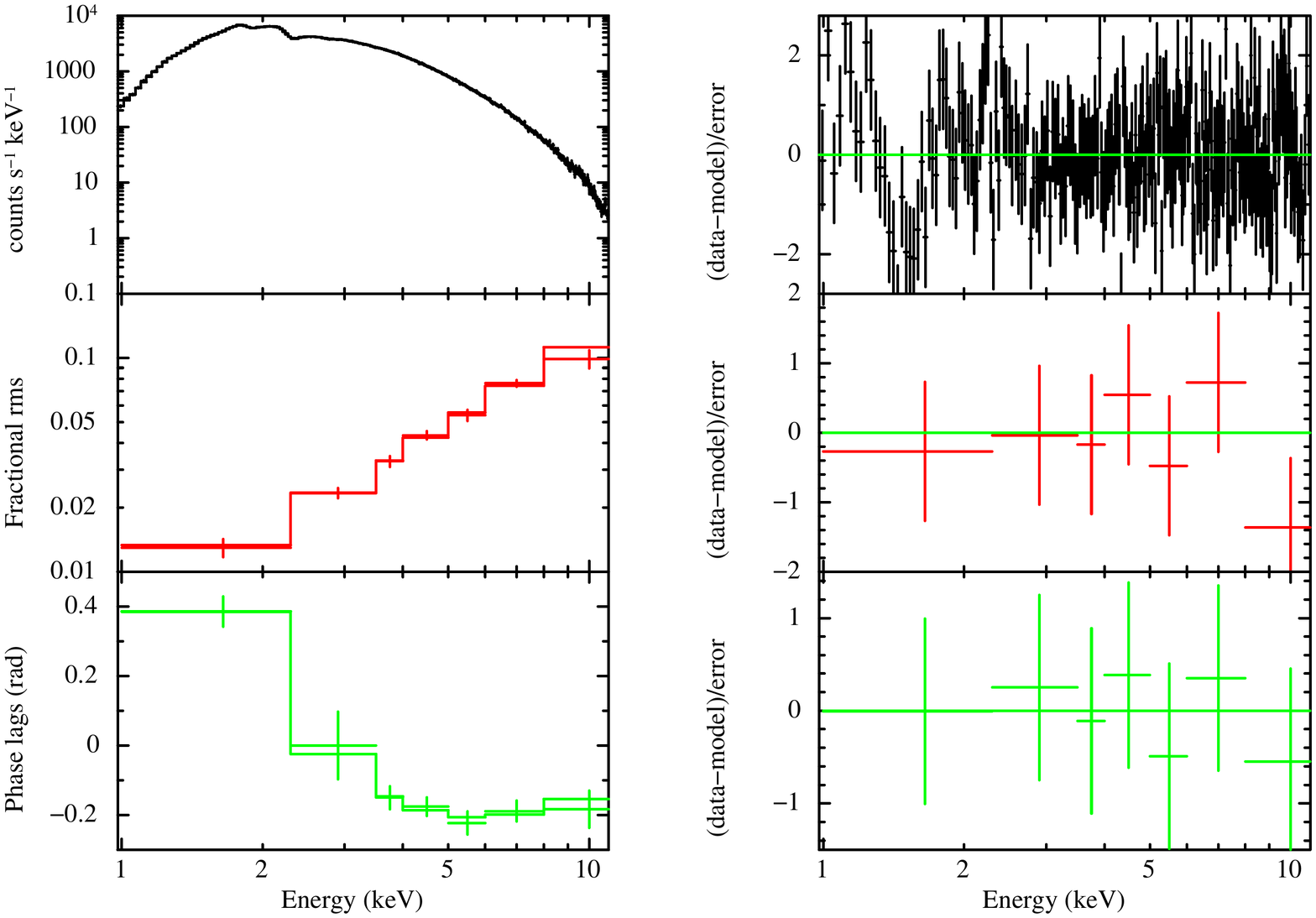}
\caption
{The same plot as shown in Figure \ref{rms_lag} at $\sim$7.0 Hz QPO frequency in MAXI~J1535$-$571.}
\label{appendix_fig3}
\end{figure*}

\begin{figure*}
\centering\includegraphics[width=0.45\textwidth,height=0.25\textheight]{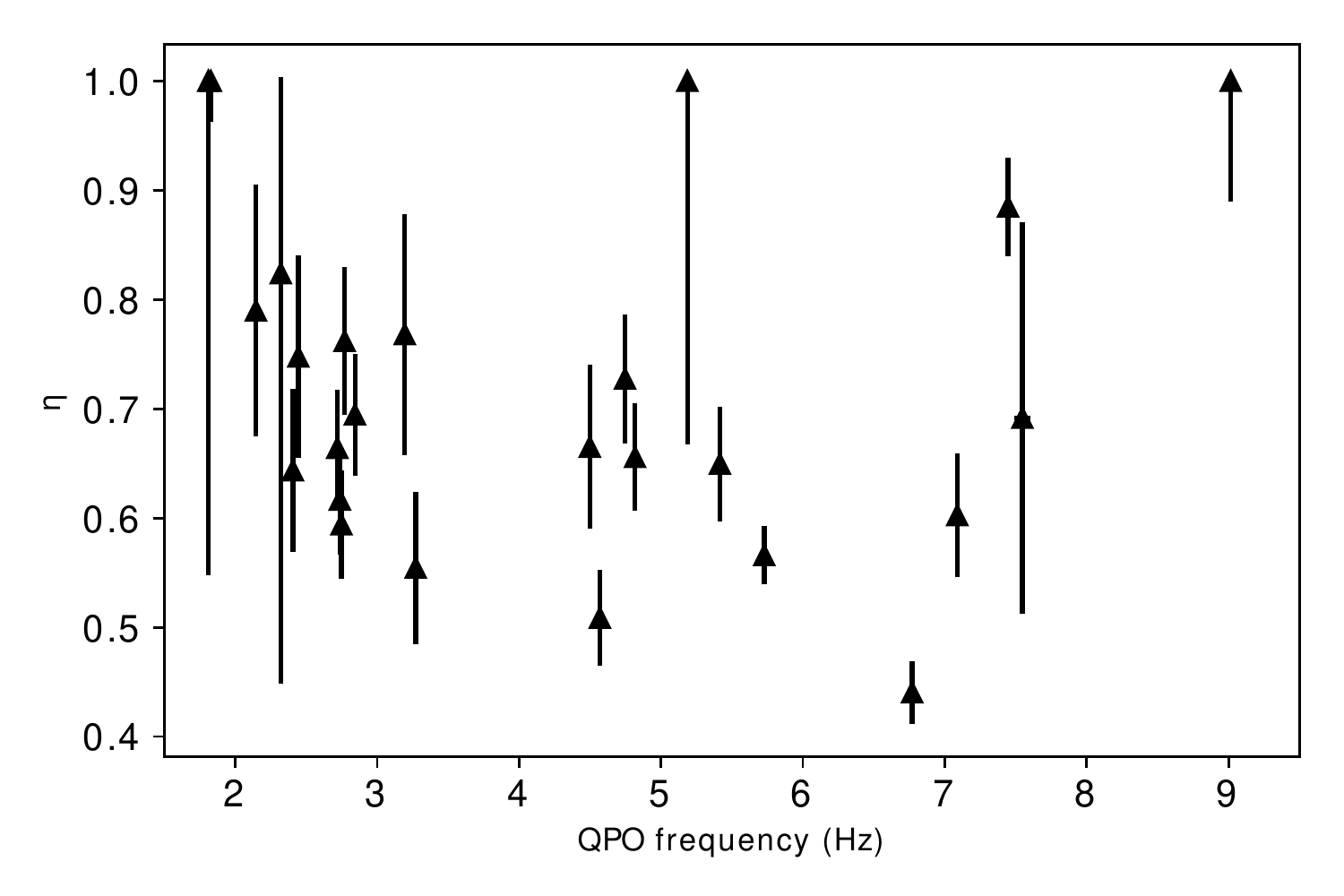}
\caption{Dependence of the $\eta$ upon QPO frequency in MAXI~J1535$-$571. The values of $\eta$ are obtained from the fits to the time-averaged spectra, the rms and phase-lag spectra of the QPO.}
\label{appendix_fig_eta_qpo}
\end{figure*}
\end{document}